\documentclass[final]{siamltex}

\usepackage{amssymb,amsmath,pstricks}
\usepackage[dvips]{epsfig}
\usepackage{graphics,graphicx,caption}

\makeatletter\@addtoreset {equation}{section}\makeatother

\def\brem{\begin{remark}}%
\def\erem{\end{remark}}%
\def\bthm{\begin{theorem}}%
\def\ethm{\end{theorem}}%
\def\blem{\begin{lemma}}%
\def\elem{\end{lemma}}%
\def\bcor{\begin{corollary}}%
\def\ecor{\end{corollary}}%
\def\bdefi{\begin{definition}}%
\def\edefi{\end{definition}}%
\def\beq{\begin{equation}}%
\def\eeq{\end{equation}}%
\def\bpf{\begin{proof}}%
\def\epf{\end{proof}}%

\newcommand{\B}{{\mathbb B}}%
\newcommand{\R}{{\mathbb R}}%
\newcommand{\C}{{\mathbb C}}%
\newcommand{\N}{{\mathbb N}}%
\newcommand{\Z}{{\mathbb Z}}%
\def\CT{{\cal T}}%
\def\CA{{\cal A}}%
\def\CE{{\cal E}}%
\def\CD{{\cal D}}%
\def\CF{{\cal F}}%
\def\CH{{\cal H}}%
\def\CP{{\cal P}}%
\def\CN{{\cal N}}%

\def\pa{{\partial}}%
\def\Lam{\varLambda}%

\newcommand{\bi}{\begin{itemize}}%
\newcommand{\ei}{\end{itemize}}%
\newcommand{\ben}{\begin{enumerate}}%
\newcommand{\een}{\end{enumerate}}%
\newcommand{\bce}{\begin{center}}%
\newcommand{\ece}{\end{center}}%
\newcommand{\barr}{\begin{array}}%
\newcommand{\earr}{\end{array}}%
\newcommand{\bpm}{\begin{pmatrix}}%
\newcommand{\epm}{\end{pmatrix}}%
\newcommand{\bspm}{\left(\begin{smallmatrix}}%
\newcommand{\espm}{\end{smallmatrix}\right)}%
\newcommand{\bgat}{\begin{gather}}%
\newcommand{\egat}{\end{gather}}%

\newcommand{\mbf}[1]{{\mathbf{#1}}}%
\newcommand{\ov}[1]{{\overline {#1}}}%
\def\chici{\chi_{\rm ci}^{(3)}}
\def\dd{\,{\rm d}}%
\def\eps{\varepsilon}%
\def\ri{{\rm i}}%
\def\Bsm{{\scriptscriptstyle\mathbb{B}}}%
\def\barl#1{\overline{#1}}
\def\hphm{\hphantom{-}}
\def\pkt{\,.\,}
\def\solleq{\stackrel{!}{=}}%
\def\wh{\widehat}%
\def\wt{\widetilde}%

\newtheorem{remark}[theorem]{Remark}

\begin{document}

\title{\bf Coupled Mode Equation Modeling for Out-of-Plane Gap Solitons in 2D Photonic Crystals}

\author{Tom\'{a}\v{s} Dohnal\footnotemark[1] \ and Willy D\"orfler\footnotemark[1]}


\date{\today}
\maketitle

\renewcommand{\thefootnote}{\fnsymbol{footnote}}

\footnotetext[1]{Institute for Applied and Numerical Mathematics, Karlsruhe
Institute of Technology,  D--76128 Karlsruhe, Germany}


\begin{abstract} Out-of-plane gap solitons in 2D photonic crystals are optical beams
localized in the plane of periodicity of the medium and delocalized in the
orthogonal direction, in which they propagate with a nonzero velocity. We study
such gap solitons as described by the Kerr nonlinear Maxwell system. Using a
model of the nonlinear polarization, which does not generate higher harmonics,
we obtain a closed curl-curl problem for the fundamental harmonic of the gap
soliton. For gap solitons with frequencies inside spectral gaps and in an
asymptotic vicinity of a gap edge we use a slowly varying envelope
approximation based on the linear Bloch waves at the edge and slowly varying
envelopes. We carry out a systematic derivation of the coupled mode equations
(CMEs) which govern the envelopes. This derivation needs to be carried out in
Bloch variables. The CMEs are a system of coupled nonlinear stationary
Schr\"odinger equations with an additional cross derivative term. Examples of
gap soliton approximations are numerically computed for a photonic crystal with
a hexagonal periodicity cell and an annulus material structure in the cell.
\end{abstract}



\pagestyle{myheadings}
\thispagestyle{plain}
\markboth{T. DOHNAL and W. D\"ORFLER}{OUT-OF-PLANE GAP SOLITONS IN 2D PHOTONIC CRYSTALS}

\section{Introduction}

Maxwell's equations for electromagnetic waves in Kerr nonlinear
dielectric materials read
\begin{subequations}\label{E:Maxwell}
 \begin{align}
 \label{E:Ampere}
   \pa_t \CD &= \nabla \times \CH,\\
 \label{E:Faraday}
   \mu_0\pa_t \CH &= -\nabla \times \CE,\\
 \label{E:Gauss}
   \nabla \cdot \CD &= 0,\\
 \label{E:Gauss_magn}
   \nabla \cdot \CH &= 0
 \end{align}
\end{subequations}
for the \emph{electric field\/} $\CE$, \emph{magnetic field\/} $\CH$,
the \emph{electric displacement field\/} $\CD$ with the
\emph{constitutive relations}
\beq\label{E:constit_rel-1}%
 \begin{split}
   \CD &= \eps_0 \left(n^2\CE+\CP_{\text{NL}}\right),\\
   \CP_{\text{NL},i}
   &= \sum_{j,l,m=1}^3 \chi_{ijlm}^{(3)}\CE_j\CE_l\CE_m
      \qquad\text{for }i=1,2,3.
 \end{split}
\eeq%
$\eps_0, \mu_0$ are the \emph{electric permittivity\/} and \emph{magnetic
permeability of vacuum}, respectively, $x\mapsto n(x)$ is the \emph{refractive
index} of the medium, and $x\mapsto\chi^{(3)}(x)$ is the \emph{cubic electric
susceptibility\/} of the medium.

We consider a 2D photonic crystal, i.e.{} we assume that the material
coefficients change periodically on a plane and are independent of the
orthogonal component on that plane. Let $a^{(1)},a^{(2)}\in\R^3$ be linearly
independent lattice vectors defining the \emph{Bravais lattice\/}
$\Lam:=\operatorname{span}_{\Z}\{a^{(1)},a^{(2)}\}$ of the crystal. Then the
required periodicity reads
\beq\label{E:constit_rel-2}%
 \begin{split}
   n(x) &= n(x+R)\in\R,\\
   \chi^{(3)}(x) &= \chi^{(3)}(x+R)\in\R^{3\times 3\times 3\times 3}
   \qquad\text{for all }x\in\R^3\text{ and all }R\in\Lam.
 \end{split}
\eeq%
Without loss of generality we assume that the crystal is homogeneous in the
$x_3$-direction, i.e.{} $a^{(1)}_3=a^{(2)}_3=0$ and $\pa_{x_3}n =
\pa_{x_3}\chi_{ijlm}^{(3)}=0$ for all $i,j,l,m$. We denote by $U$ the
\emph{Wigner--Seitz cell\/} corresponding to the Bravais lattice. We use
$b^{(1)}, b^{(2)}$ to denote the pair of vectors satisfying $a^{(i)}\cdot
b^{(j)}=2\pi \delta_{i,j}$ for $i,j\in \{1,2\}$, and let the \emph{reciprocal
lattice\/} be $\Lam^*:=\operatorname{span}_{\Z}\{b^{(1)},b^{(2)}\}$. $\B$
denotes the first \emph{Brillouin zone}, i.e.{} the Wigner--Seitz cell of the
reciprocal lattice.

Note that from the relations in \eqref{E:constit_rel-1} it is clear that we are
neglecting losses, material dispersion as well as higher order nonlinearities
and assuming that the third order nonlinear response of the medium is
instantaneous.

We will consider monochromatic waves propagating in the $x_3$-direction, i.e.{}
waves propagating \emph{out of the plane of periodicity} of the 2D crystal, and
use the ansatz
\beq\label{E:field_form}%
  (\CE,\CH,\CD)(x,t)= e^{\ri(\kappa x_3-\omega t)}
     (E,H,D)(x_1,x_2;\omega) + \text{c.c.},
\eeq%
where $\kappa \in \R$ and c.c.{} denotes the complex conjugate of the first
term on the right. The ansatz \eqref{E:field_form} contains no higher
harmonics, which is valid if the above form of $\CP_{\text{NL}}$ is replaced by
a time averaged one, see below. Alternatively, a physical justification of
neglecting higher harmonics is based on the lack of phase matching and absorption.

Note that for the field \eqref{E:field_form} the divergence free conditions
\eqref{E:Gauss} and \eqref{E:Gauss_magn} are automatically satisfied provided
$\omega \neq 0$ since the spatially dependent parts
\[
   \big(\hat{\CE},\hat{\CH},\hat{\CD}\big)(x;\omega)
   :=e^{\ri\kappa x_3}\big(E,H,D\big)(x_1,x_2;\omega)
\]
satisfy
\beq\label{E:solenoid}%
   \hat{\CD}=\frac{\ri}{\omega}\nabla\times\hat{\CH} \quad
   \text{and} \quad \mu_0\hat{\CH}=-\frac{\ri}{\omega}\nabla\times\hat{\CE},
\eeq%
and thus $\nabla\cdot\hat{\CD} = \nabla\cdot\hat{\CH}=0$. Since our analysis
below is for gap solitons with $\omega$ close to a band edge, the condition
$\omega\neq 0$ is for us restrictive only when $\omega=0$ is in a gap and lies
near a band edge. Note also that even if higher harmonics are accounted for,
the divergence free conditions are still satisfied for $\omega \neq 0$ as
\eqref{E:solenoid} then holds for each generated harmonic. Clearly, only odd,
i.e., $(2n+1)$-th, $n\in\Z$, harmonics are generated.

We will assume a centrosymmetric and isotropic $\chi^{(3)}$-tensor, which leads
to the simplification
\begin{align*}
   \CP_{\text{NL}}= \chici (\CE \cdot \CE) \CE,
\end{align*}%
where $\chici := \chi^{(3)}_{1111} = \chi^{(3)}_{2222} = \chi^{(3)}_{3333}$ for
$\chici:(x_1,x_2)\in \R^2 \rightarrow \R$, see \cite[Sec.~2d]{MN04}. Inserting
the ansatz \eqref{E:field_form} in the nonlinearity $\CP_{\text{NL}}$ clearly
generates the harmonics $e^{\pm 3\ri(\kappa x_3-\omega t)}$. These are,
however, typically neglected based on the physical arguments that the
fundamental harmonics $e^{\pm \ri(\kappa x_3-\omega t)}$ and the higher
harmonics are not phase matched and that at the higher values of frequency
(i.e. at $\pm 3\omega$) material absorption is usually large preventing the
generation of significant fields at these frequencies, see e.g.{} \cite{BS01}.
Considering only the fundamental harmonics, the nonlinear polarization for the
ansatz \eqref{E:field_form} becomes
\beq\label{E:PNL_ansatz}%
   \CP_{\text{NL}}
   =\chici \big(2|E|^2\,E+E\cdot E \,\barl{E}\big)
    e^{\ri(\kappa x_3-\omega t)} + \text{c.c.}\,,
\eeq%
i.e.{}
\beq\label{E:PNL_ans_full}
 \CP_{\text{NL}}
 =\chici \bspm (3|E_1|^2+2|E_2|^2+2|E_3|^2)E_1+(E_2^2+E_3^2)\bar{E}_1\\
  (2|E_1|^2+3|E_2|^2+2|E_3|^2)E_2+(E_1^2+E_3^2)\bar{E}_2\\
  (2|E_1|^2+2|E_2|^2+3|E_3|^2)E_3+(E_1^2+E_2^2)\bar{E}_3
  \espm e^{\ri(\kappa  x_3-\omega t)} + \text{c.c.}.
\eeq%
Another widely used model for the nonlinear polarization is
\begin{align*}
   \CP_{\text{NL}}= \chici [ \CE \cdot \CE ]^{\rm av} \CE,
\end{align*}%
where $[f]^{\rm av}$ denotes the time average of $f$ over the period of $f$,
i.e.{} over $t\in [0,\pi/\omega]$ for $f=\CE \cdot \CE$, cf.{}
\cite{Stuart_93,Sutherland03}. The averaging generates no higher harmonics so
that in this model \eqref{E:PNL_ansatz} is exact. Note that the Kerr nonlinear
problem including all higher harmonics has been recently considered for a 1D
periodic structure in \cite{PSW11}.

In the following we rescale the frequency by defining
$$
   \wt{\omega}:=\frac{\omega}{c}
$$
but drop the tilde again for better readability. For convenience we will denote
the square of the refractive index by
$$
   \eta(x):=n^2(x)\qquad\text{for all }x\in\R^3.
$$

With the ansatz \eqref{E:field_form} equations \eqref{E:Ampere} and
\eqref{E:Faraday} become
\begin{subequations}\label{E:Maxw_1st_harm}
\begin{align}
  -\ri c \omega D  &= \nabla \times H + \ri \bspm 0\\0\\\kappa \espm\times H,\\
  \ri c \omega \mu_0 H &= \nabla \times E + \ri \bspm 0\\0\\\kappa \espm\times E.
\end{align}
\end{subequations}
Since all our functions are independent of $x_3$, we let from now on
$x=(x_1,x_2)\in\R^2$. Using the fact that $E$ depends only on $x_1$ and $x_2$,
a second order formulation of \eqref{E:Maxw_1st_harm} reads
\beq\label{E:NL_Maxw}%
 \left(L-\omega^2\eta\right)E = \omega^2 P_{\text{NL}},
\eeq%
where
\beq\label{E:L_op}%
   L E := \nabla\times \nabla \times E + \ri \kappa
      \bspm\pa_{x_1}E_3\\\pa_{x_2}E_3\\\pa_{x_1}E_1+\pa_{x_2}E_2\espm
      +\kappa^2 \bspm E_1\\E_2\\0\espm,
\eeq%
and
\[
   P_{\text{NL}}
   = \chici \big(2|E|^2\,E+E\cdot E \,\barl{E}\big).
\]%
Having determined $E$, the magnetic field can be recovered by
\[
   H = -\tfrac{\ri}{\omega \mu_0}\left(\nabla \times E
      + \ri \bspm 0\\0\\\kappa\espm\times E\right).
\]%

Based on the analogy with the periodic nonlinear Schr\"odinger equation
\cite{Pankov05}, equation \eqref{E:NL_Maxw} is expected to have localized
$H(\text{curl},\R^2)$-solutions $E$ for any $\omega$ in a spectral gap of the
linear problem $Lu=\omega^2 \eta u$. Such solutions are called \textit{gap
solitons}. The aim of this paper is to provide an approximation of gap solitons
$E$ of \eqref{E:NL_Maxw} for $\omega$ in an $\eps^2$-vicinity ($0<\eps\ll 1$)
of a gap edge using a slowly varying envelope approximation. As we show,
envelopes of such gap solitons satisfy a system of nonlinear constant
coefficient equations, so called \textit{coupled mode equations} (CMEs) posed
in the slow variables $y=\eps x$. The CMEs can be numerically solved with less
effort than the nonlinear Maxwell system \eqref{E:NL_Maxw} in the variable $x$.
An asymptotic approximation of a gap soliton of \eqref{E:NL_Maxw} near a gap
edge is then the sum of linear Bloch waves at the edge, modulated by the
corresponding envelopes.

Asymptotic approximations via CMEs have been analyzed for gap solitons of the
stationary periodic nonlinear Schr\"odinger equation in 1D \cite{PSn07} as well
as in 2D \cite{DPS09,DU09,DU_err11}. In these works the approximation via CMEs
was also rigorously justified using Lyapunov--Schmidt reductions. Gap solitons
of the nonlinear Maxwell's equations have been approximated by CMEs in the case
of 1D photonic crystals with a small (infinitesimal) contrast in the
periodicity \cite{GWH01,PSn07,PSW11}, where \cite{GWH01} considers gap solitons
modulated also in time. To our knowledge the problem of a systematic CME
approximation of gap solitons of nonlinear Maxwell's equations describing 2D or
3D photonic crystals does not appear in the literature. Although CMEs have been
formally derived for pulses in Maxwell's equations with a 2D periodic medium of
small contrast \cite{AJ98,AP05,DA05}, these pulses cannot be true gap solitons
because in 2D and 3D a large enough contrast is necessary for the opening of
spectral gaps. In this paper we consider a 2D photonic crystal with a finite
contrast in the periodicity. For our examples we use a photonic crystal which
has several spectral gaps \cite{AM03}.

Besides the above cited works on coupled mode modeling of gap solitons there
are a number of papers on the slowly varying envelope approximation of
nonlinear pulses in periodic structures with the pulse frequency lying within
the spectral bands. The envelope in this case can be typically modeled by the
time dependent nonlinear Schr\"odinger equation and the approximation holds on
large but finite time intervals \cite{BS01,BabinFigotin:2005,BSTU06}.

The rest of the paper is organized as follows. In Section \ref{S:band_str} we
study the linear band structure $\omega_n(k)$ of \eqref{E:NL_Maxw} (with
$\chici=0$) and obtain thus the linear spectrum of the problem. We also discuss
possible symmetries in the band structure and among the corresponding Bloch
waves. An example of a photonic crystal from \cite{AM03} is then provided, for
which the band structure is numerically computed and three band gaps are
observed on the positive half axis $\omega>0$. In Section \ref{S:CME_deriv} we
present a slowly varying envelope approximation of gap solitons of
\eqref{E:NL_Maxw} for $\omega$ in the vicinity of a spectral edge and carry out
a systematic formal derivation of CMEs describing the envelopes. Next, examples
of CMEs are presented for the concrete photonic crystal given in Section
\ref{S:band_str} as well as for other theoretical situations. Here the
symmetries in the band structure and among the Bloch waves play an important
role in determining properties of the CME coefficients. In Section
\ref{S:numerics} we plot the approximation of two gap solitons in the chosen
photonic crystal. The approximation requires computing the Bloch waves at the
edge and solving the corresponding CMEs.

\section{Linear Band Structure}\label{S:band_str}

\subsection{The periodic eigenvalue problem}\label{S:evp}

We study first the linear problem
\beq\label{E:lin_Maxw_2nd}
   Lu=\omega^2 \eta u\qquad\text{on }\R^2
\eeq
and define the band structure as well as the linear Bloch waves.

By the Bloch--Floquet theory, see \cite{Kuchment_1993} or
\cite[Ch.~3]{DLPSW_2011}, solution modes of \eqref{E:lin_Maxw_2nd} are given by
the \emph{Bloch waves\/} $u_n(k;\pkt)$ for $n\in\N$ that satisfy
\begin{equation}\label{E:omega_nk_def}%
\begin{split}
   Lu_n(k;\pkt)&=\omega_n(k)^2\eta u_n(k;\pkt),\\
   u_n(k;\pkt+R)&=u_n(k;\pkt)e^{\ri k\cdot R}
   \qquad\quad\text{for all }R\in\Lam,
\end{split}
\end{equation}
where $k=(k_1,k_2)$ sweeps the first Brillouin zone $\B\subset \R^2$.

It is well-known that $L$ is self-adjoint and has a compact inverse and that
there thus exists a sequence of eigenvalues $\{\omega_n\}_{n\ge1}$ with
$\lim_{n\to\infty}\omega_n=\infty$ and each eigenspace is of finite dimension.
These eigenvalues are nonnegative and we use the natural ordering
$\omega_{n-1}\le\omega_n$ for $n\ge1$. The mapping $k\mapsto\omega_n(k)$ is
called the \emph{$n$-th band} of the spectral problem \eqref{E:omega_nk_def}.
Of course, \eqref{E:omega_nk_def} allows also non-positive bands $-\omega_n$.
These are typically labeled via $\omega_{-n}=-\omega_n$ and will play no role
in our analysis. We therefore restrict ourselves to $\omega_n \geq 0$ for $n
\in \N$. The Bloch waves in \eqref{E:omega_nk_def} can be written in the form
\[%
   u_n(k;x) = p_n(k;x)e^{\ri k\cdot x},
\]%
where the $p_n$ are $\Lambda$-periodic in $x$, i.e.{} $p_n(k;x+R)=p_n(k;x)$
for all $x\in U$, $R\in\Lam$. These satisfy the eigenvalue problem
\beq\label{E:shifted}%
\begin{split}
   \left(\wt{L}(k) -\omega_n^2(k)\eta(x)\right) p_n(k;x)
   &=0\qquad\qquad\quad\text{for all }x\in U,\\
   p_n(k;x+R)&=p_n(k;x)\qquad\text{for all }x\in\partial U
      \text{ and all }R\in\Lam,
\end{split}
\eeq%
with
$$
   \wt{L}(k) p_n(k;x)
   = (\nabla + \ri k')\times(\nabla + \ri k')\times p_n(k;x),
$$
where $k=(k_1,k_2)\in\B$, $k'=(k_1,k_2,\kappa)^T$. Since $p_n$ is
$x_3$-independent, $\wt{L}(k)$ can be written as
\[
 \wt{L}(k) =
 \bspm\kappa ^2-(\pa_{x_2}+\ri k_2)^2 & (\pa_{x_1}+\ri k_1)(\pa_{x_2}+\ri k_2)
     & \ri \kappa (\pa_{x_1}+\ri k_1)\\
 (\pa_{x_1}+\ri k_1)(\pa_{x_2}+\ri k_2) & \kappa ^2-(\pa_{x_1}+\ri k_1)^2
     & \ri \kappa (\pa_{x_2}+\ri k_2)\\
 \ri \kappa (\pa_{x_1}+\ri k_1) & \ri \kappa (\pa_{x_2}+\ri k_2)
     & -(\pa_{x_1}+\ri k_1)^2-(\pa_{x_2}+\ri k_2)^2
 \espm.
\]%
In the variable $k$ the Bloch waves $u_n$ and the eigenvalues $\omega_n$ are
easily proved to fulfill
\beq\label{E:k-per}%
   \omega_n(k) = \omega_n(k+K),
   \quad p_n(k+K;x)=p_n(k;x)e^{-\ri K\cdot x}
   \qquad \text{for all }x\in U,\,K\in\Lam^*.
\eeq%
Due to the self-adjoint nature of $\wt{L}(k)$ we can normalize the Bloch
functions via
\beq\label{E:normalize_Bloch}%
   \left\langle p_n(k;\pkt),\eta p_m(k;\pkt)\right\rangle
   = \delta_{n,m},
\eeq%
where $\langle f,g \rangle = \langle f,g \rangle_{L^2(U)^3} = \int_{U}
f(x)\cdot\barl{g}(x)\dd x$ for $f,g:\R^2\rightarrow\C^3$.

For purposes of the later asymptotic analysis of gap solitons we also present
calculations of first and second order derivatives of the bands at extremal
points. Suppose the band $\omega_{n_*}$ has an extremum at $k=k_*\in \B$ and
denote $\omega_*:=\omega_{n_*}(k_*)$. By direct differentiation of
\eqref{E:shifted} we see that the ``generalized Bloch functions''
$\pa_{k_j}p_{n_*}$, for $j \in \{1,2\}$, are solutions of the system
\beq\label{E:deriv_eq}%
 \left(\wt{L}(k_*)-\omega_*^2\eta\right)\pa_{k_j}p_{n_*}(k_*;\pkt)
 =- \pa_{k_j}\wt{L}(k_*) p_{n_*}(k_*;\pkt).
\eeq%
Applying the differentiation $\pa_{k_i,k_j}^2$, for $i,j\in\{1,2\}$, to
\eqref{E:shifted} and evaluation at $n=n_*$, $k=k_*$ yields
\begin{align*}
    &\left(\wt{L}(k_*)-\omega_*^2\eta(x)\right)
     \pa_{k_i,k_j}^2p_{n_*}(k_*;x)\\
    &\qquad = 2\omega_*\eta(x)\pa_{k_i,k_j}^2
      \omega_{n_*}(k_*)p_{n_*}(k_*;x)
       -\pa_{k_i,k_j}^2\wt{L}(k_*)p_{n_*}(k_*;x)\\
    &\qquad\quad{}-\pa_{k_i}\wt{L}(k_*)\pa_{k_j}p_{n_*}(k_*;x)
       -\pa_{k_j}\wt{L}(k_*)\pa_{k_i}p_{n_*}(k_*;x).
\end{align*}
Necessarily, due to the Fredholm alternative, the right hand side is
$L^2$-orthogonal to $p_{n_*}(k_*;\pkt)$, which lies in the kernel of
$\wt{L}(k_*)-\omega_*^2\eta$ with periodic boundary conditions on $U$. This
yields the formula
\begin{align}\label{E:om_deriv_2}\notag%
   &\left(\pa_k^2\omega_{n_*}(k_*)\right)_{i,j}
    =\pa_{k_i,k_j}^2\omega_{n_*}(k_*)\\
   &\quad=\frac{1}{2\omega_*}\left\langle \pa_{k_i,k_j}^2
     \wt{L}(k_*)p_{n_*}(k_*;\pkt)
     +\pa_{k_i}\wt{L}(k_*)\pa_{k_j}p_{n_*}(k_*;\pkt)
     +\pa_{k_j}\wt{L}(k_*)\pa_{k_i}p_{n_*}(k_*;\pkt),
     p_{n_*}(k_*,\cdot)\right\rangle.
\end{align}%
A straightforward differentiation of $\wt{L}(k)$ yields
\begin{align*}%
 \pa_{k_1}\wt{L}(k_*)
 &=\bspm
 0 & \ri (\pa_{x_2}+\ri k_{*,2}) & -\kappa \\
 \ri (\pa_{x_2}+\ri k_{*,2}) & -2\ri (\pa_{x_1}+\ri k_{*,1}) & 0\\
 -\kappa & 0 & -2\ri(\pa_{x_1}+\ri k_{*,1})
 \espm,\\
 \pa_{k_2}\wt{L}(k_*)
 &=\bspm
 -2\ri (\pa_{x_2}+\ri k_{*,2}) & \ri (\pa_{x_1}+\ri k_{*,1}) & 0\\
 \ri (\pa_{x_1}+\ri k_{*,1}) & 0& -\kappa \\
 0& -\kappa & -2\ri(\pa_{x_2}+\ri k_{*,2})
 \espm,
\end{align*}
\beq\label{E:Ltil_2nd_der}
 \pa_{k_1}^2\wt{L}\equiv \bspm 0&0&0\\0&2&0\\0&0&2\espm,
 \quad \pa_{k_2}^2\wt{L}\equiv \bspm 2&0&0\\0&0&0\\0&0&2\espm,
 \quad \text{and} \quad \pa_{k_1,k_2}^2\wt{L}\equiv
 \bspm \hphm 0&-1&\hphm 0\\-1&\hphm 0&\hphm 0\\\hphm 0&\hphm 0&\hphm 0\espm,
\eeq%
where $k_{*,j}$, for $j\in\{1,2,3\}$, is the $j$-th component of $k_*$. With
these the explicit forms of \eqref{E:om_deriv_2} read
\beq\label{E:om_2deriv_k1}%
 \pa_{k_1}^2\omega_{n_*}(k_*) =
 \frac{1}{\omega_*} \left\langle \bspm
 \ri(\pa_{x_2}+\ri k_{*,2})\pa_{k_1}p_{n_*,2}(k_*;\pkt)
 -\kappa \pa_{k_1}p_{n_*,3}(k_*;\pkt)\\
 \ri(\pa_{x_2}+\ri k_{*,2})\pa_{k_1}p_{n_*,1}(k_*;\pkt)
 -2\ri(\pa_{x_1}+\ri k_{*,1})
 \pa_{k_1}p_{n_*,2}(k_*;\pkt)+p_{n_*,2}(k_*;\pkt)\\
 -2\ri(\pa_{x_1}+\ri k_{*,1})\pa_{k_1}p_{n_*,3}(k_*;\pkt)
 -\kappa \pa_{k_1}p_{n_*,1}(k_*;\pkt)+p_{n_*,3}(k_*;\pkt)
 \espm,p_{n_*}(k_*;\pkt)
 \right\rangle,
\eeq%
\beq\label{E:om_2deriv_k2}%
 \pa_{k_2}^2\omega_{n_*}(k_*)
 = \frac{1}{\omega_*}\left\langle
 \bspm
 -2\ri(\pa_{x_2}+\ri k_{*,2})\pa_{k_2}p_{n_*,1}(k_*;\pkt)
 +\ri(\pa_{x_1}+\ri k_{*,1})
 \pa_{k_2}p_{n_*,2}(k_*;\pkt)+p_{n_*,1}(k_*;\pkt)\\
 \ri(\pa_{x_1}+\ri k_{*,1})\pa_{k_2}p_{n_*,1}(k_*;\pkt)
 -\kappa \pa_{k_2}p_{n_*,3}(k_*;\pkt)\\
 -\kappa \pa_{k_2}p_{n_*,2}(k_*;\pkt)-2\ri(\pa_{x_2}+\ri k_{*,2})
 \pa_{k_2}p_{n_*,3}(k_*;\pkt)+p_{n_*,3}(k_*;\pkt)
 \espm,p_{n_*}(k_*;\pkt)
 \right\rangle,
\eeq%
and
\begin{align}\label{E:om_deriv_k12}%
 \barr{rl}
 \pa_{k_1,k_2}^2\omega_{n_*}(k_*) = &\frac{1}{2\omega_*}
 \left\langle
 \bspm
 -2\ri(\pa_{x_2}+\ri k_{*,2})\pa_{k_1}p_{n_*,1}(k_*;\pkt)
 +\ri(\pa_{x_1}+\ri k_{*,1})\pa_{k_1}p_{n_*,2}(k_*;\pkt)+\ri(\pa_{x_2}
 +\ri k_{*,2})\pa_{k_2}p_{n_*,2}(k_*;\pkt)\\
 \ri(\pa_{x_1}+\ri k_{*,1})\pa_{k_1}p_{n_*,1}(k_*;\pkt)+\ri(\pa_{x_2}
 +\ri k_{*,2})\pa_{k_2}p_{n_*,1}(k_*;\pkt)-2\ri(\pa_{x_1}+\ri k_{*,1})
 \pa_{k_2}p_{n_*,2}(k_*;\pkt)\\
 -2\ri(\pa_{x_2}+\ri k_{*,2})\pa_{k_1}p_{n_*,3}(k_*;\pkt)-2\ri(\pa_{x_1}
 +\ri k_{*,1})\pa_{k_2}p_{n_*,3}(k_*;\pkt)
 \espm
 \right.\\
 & \qquad \left. {}+
 \bspm
 -\kappa \pa_{k_2}p_{n_*,3}(k_*;\pkt)-p_{n_*,2}(k_*;\pkt)\\
 -\kappa \pa_{k_1}p_{n_*,3}(k_*;\pkt)-p_{n_*,1}(k_*;\pkt)\\
 -\kappa \big(\pa_{k_1}p_{n_*,2}(k_*;\pkt)+\pa_{k_2}p_{n_*,1}(k_*;\pkt)\big)
 \espm, p_{n_*}(k_*;\pkt)
 \right\rangle.
 \earr
\end{align}%

\subsection{Symmetries of the Band Structure and the Bloch waves}

Symmetries in the refractive index function $\eta$ yield symmetries in the band
structure and among Bloch waves. We restrict our attention to the cases of
discrete rotational and axial reflection symmetry, which are relevant for the
example we present below. The results of this section will be important when
determining properties of the coefficients of coupled mode equations in Section
\ref{S:CME_examples}.

\subsubsection{Rotational symmetry}\label{S:rot_sym}
Assume that the photonic crystal satisfies the rotational symmetry
\beq\label{E:material_rot_sym}%
   \eta(x) = \eta(r_\alpha(x)) \qquad \text{for all }x\in \R^2
\eeq%
for some $\alpha \in (-\pi,\pi]$ with the rotation $r_\alpha$ defined
by
\[
  r_\alpha(x) =
  \bspm
  \cos(\alpha)x_1 -\sin(\alpha) x_2\\
  \sin(\alpha)x_1 +\cos(\alpha) x_2
  \espm
\]%
Below we use the notation $r_\alpha(v) = (\cos(\alpha)v_1 -\sin(\alpha) v_2,
\sin(\alpha)v_1 +\cos(\alpha) v_2)^T$ if $v$ is a two dimensional vector $v\in
\C^2$ and $r_\alpha(v) = (\cos(\alpha)v_1 -\sin(\alpha) v_2, \sin(\alpha)v_1
+\cos(\alpha) v_2, v_3)^T$ if $v$ is a three dimensional vector $v\in \C^3$.

The symmetry \eqref{E:material_rot_sym} implies a symmetry of the Rayleigh
quotient corresponding to the eigenvalue problem \eqref{E:shifted} and thus a
symmetry of the band structure. In detail, for $k\in \B$ we have
\[%
 \omega_n^2(k)=\min_{\substack{V\subset H^{\text{curl}}_{\text{per}}(U) \\
 \dim V=n}} \quad \max_{w\in V,\,w\neq 0}
 \frac{\int_U |(\nabla +\ri k')\times w(x)|^2\dd x}%
    {\int_U\eta(x) |w(x)|^2\dd x},
\]%
and the corresponding extremal point is $p_n(k;\pkt)$. Due to the
relation
$$
 \left((\nabla+\ri r_\alpha(k'))\times f\right)(r_{\alpha}(x))
 = r_{\alpha}\left[(\nabla +\ri k')\times r_{-\alpha}
 \left(f(r_{\alpha}(x))\right)\right]
 \quad \text{for all smooth }f:\R^2\to\R^3
$$
we get
$$
 \int_U |(\nabla +\ri r_\alpha(k'))\times w(x)|^2\dd x
 = \int_U |(\nabla +\ri k')\times r_{-\alpha}
 \left(w(r_{\alpha}(x))\right)|^2\dd x,
$$
and symmetry \eqref{E:material_rot_sym} yields
$$
 \int_U \eta(x) |w(x)|^2\dd x
 = \int_U \eta(x) |r_{-\alpha}(w(r_{\alpha}(x))) |^2\dd x.
$$
As a result we obtain that
\beq\label{E:rot_sym_bands}%
  \omega_n(k) = \omega_n(r_\alpha(k)) \qquad \text{for all }n \in \N
    \text{ and all }k\in\B.
\eeq%
If $\omega_n(k)$ has geometric multiplicity one as an eigenvalue of
\eqref{E:shifted}, we have also a symmetry of the corresponding Bloch
functions, namely
\beq\label{E:rot_sym_Bloch}%
  p_n(r_\alpha(k);x) = e^{\ri a} r_{-\alpha}
  \left(p_n(k;r_{\alpha}(x))\right)
  \qquad \text{for all } n \in \N \text{ and some } a=a(n)\in \R .
\eeq%
Note that a renormalization of $p_n(r_\alpha(k);x)$, in order to obtain $a=0$
in \eqref{E:rot_sym_Bloch}, is in general impossible when $r_\alpha(k)\doteq
k$, where $k\doteq l$ reads ``$k$ congruent to $l$'' and means $k=l+K$ for some
$K\in\Lam^*$. This is because in this case $p_n(r_\alpha(k);x)$ and $p_n(k;x)$
are related by \eqref{E:k-per} and a renormalization of the left hand side of
\eqref{E:rot_sym_Bloch} would affect the right hand side in the same way. When
$r_\alpha(k)$ is not congruent to $k$, e.g.{} when $k\in
\operatorname{int}(\B)\setminus \{0\}$, then one can set $a=0$ in
\eqref{E:rot_sym_Bloch}.

>From the symmetry \eqref{E:rot_sym_bands} we can deduce a symmetry of the
second derivatives of $\omega_n$. Using the identity $\partial_{k}\omega_n(k)
=\partial_{k}(\omega_n(r_\alpha(k)))
=(r_\alpha)^T(\partial_{k}\omega_n)(r_\alpha(k))$, we get by further
differentiation
\beq\label{E:der2_rot_sym}%
 \begin{pmatrix}
  \partial_{k_1}^2\omega_n(r_\alpha(k))\\
  \partial_{k_2}^2\omega_n(r_\alpha(k))\\
  \partial_{k_1,k_2}^2\omega_n(r_\alpha(k))
 \end{pmatrix}
 =\begin{pmatrix}
    \cos^2(\alpha) & \sin^2(\alpha) & -\sin(2\alpha)\\
    \sin^2(\alpha)&\cos^2(\alpha)&\sin(2\alpha)\\
    \tfrac{1}{2}\sin(2\alpha)&-\tfrac{1}{2}\sin(2\alpha)&\cos(2\alpha)
  \end{pmatrix}
  \begin{pmatrix}\partial_{k_1}^2\omega_n(k)\\
   \partial_{k_2}^2\omega_n(k)\\
   \partial_{k_1,k_2}^2\omega_n(k)
  \end{pmatrix}
\eeq%
for all $k\in \B$ and $n\in \N$.

\subsubsection{Reflection symmetry}\label{S:refl_sym}
If the photonic crystal satisfies the reflection symmetry
\beq\label{E:material_refl_sym}%
  \eta(x) = \eta(S_1(x)) \qquad \text{for all }x\in \R^2,
  \text{ where } S_1(x) = (-x_1,x_2)^T,
\eeq%
then similarly to Section \ref{S:rot_sym} we have
\beq\label{E:refl_sym_bands}%
  \omega_n(k) = \omega_n(-k_1,k_2) \qquad \text{for all } k\in \B
  \text{ and } n \in \N.
\eeq%
Again, if $\omega_n(k)$ has geometric multiplicity one as an eigenvalue of
\eqref{E:shifted}, then
\beq\label{E:refl_sym_Bloch}%
  p_n(S_1(k);x) = e^{\ri a} S_1\left(p_n(k;S_1(x))\right)
  \qquad\text{for all } n \in \N \text{ and some } a=a(n)\in \R,
\eeq%
where $S_1(v)=(-v_1,v_2,v_3)^T$ for $v\in \C^3$. Just as above, unless $k
\doteq S_1(k)$, we can set $a = 0$ in \eqref{E:refl_sym_Bloch}. The symmetry
\eqref{E:refl_sym_bands} implies
\beq\label{E:der2_refl_sym}%
 \begin{split}
  &\partial_{k_1}^2\omega_n(k)=(\partial_{k_1}^2\omega_n)(-k_1,k_2),
   \quad\partial_{k_2}^2\omega_n(k)=(\partial_{k_2}^2\omega_n)(-k_1,k_2),\\
  &\partial_{k_1,k_2}^2\omega_n(k)
   =-(\partial_{k_1,k_2}^2\omega_n)(-k_1,k_2)
 \end{split}
\eeq%
for all $k\in \B$ and $n\in \N$.

An analogous discussion, of course, applies for the reflection symmetry
$\eta(x) = \eta(S_2(x))$ for all $x\in \R^2$, where $S_2(x) = (x_1,-x_2)^T$.
One the obtains
\beq\label{E:der2_refl_symB}%
 \begin{split}
  &\partial_{k_1}^2\omega_n(k)=(\partial_{k_1}^2\omega_n)(k_1,-k_2),
   \quad \partial_{k_2}^2\omega_n(k)=(\partial_{k_2}^2\omega_n)(k_1,-k_2),\\
  &\partial_{k_1,k_2}^2\omega_n(k)
   =-(\partial_{k_1,k_2}^2\omega_n)(k_1,-k_2)
 \end{split}
\eeq%
for all $k\in \B$ and $n\in \N$ and if $\omega_n(k)$ has geometric multiplicity
one as an eigenvalue of \eqref{E:shifted}, then
\beq\label{E:refl2_sym_Bloch}%
  p_n(S_2(k);x) = e^{\ri a} S_2\left(p_n(k;S_2(x))\right)
  \quad\text{for all } n \in \N \text{ and some } a=a(n)\in \R .
\eeq%

\subsubsection{Combination of rotational and reflection symmetries}%
If both the reflection symmetry \eqref{E:material_refl_sym} and the rotational
symmetry \eqref{E:material_rot_sym} for some $\alpha \in (-\pi,-\pi]$,
$|\alpha|\neq \pi/2$, hold, then for $k$ along the rays with angles
$\pi/2-\alpha/2$ and $-(\pi/2+\alpha/2)$ the mixed derivative
$\partial_{k_1,k_2}^2\omega_n(k)$ can be expressed in terms of
$\partial_{k_1}^2\omega_n(k)$ and $\partial_{k_2}^2\omega_n(k)$. This is
because for $k$ along these rays we have $(-k_1,k_2)=r_{\alpha}(k)$ or
$(k_1,-k_2)=r_{\alpha}(k)$, so that both \eqref{E:der2_rot_sym} and
\eqref{E:der2_refl_sym} or \eqref{E:der2_refl_symB} apply. In detail, suppose
$$
   (-k_1,k_2)=r_{\alpha}(k), \text{ i.e.{} } k
   =|k|e^{\ri (\pi/2-\alpha/2)} \quad\text{or}\quad k
   =|k|e^{-\ri (\pi/2+\alpha/2)}=-|k|e^{\ri (\pi/2-\alpha/2)}.
$$
Then it follows that
\[
   \partial_{k_1}^2\omega_n(k)
   =(\partial_{k_1}^2\omega_n)(-k_1,k_2)
   =\cos^2(\alpha)\partial_{k_1}^2\omega_n(k)
    -\sin(2\alpha)\partial_{k_1,k_2}^2\omega_n(k)
    +\sin^2(\alpha)\partial_{k_2}^2\omega_n(k),
\]%
where the first equality holds due to \eqref{E:der2_refl_sym} and the second
due to \eqref{E:der2_rot_sym}. As a result, for $\alpha\in (-\pi,\pi]$,
$|\alpha| \neq \pi/2$, and $k=\pm |k|e^{\ri (\pi/2-\alpha/2)}$ we get
\beq\label{E:mix_der_slave}%
  \partial_{k_1,k_2}^2\omega_n(k)
  =\frac{1}{2}\tan(\alpha)\left(\partial_{k_2}^2\omega_n(k)
   -\partial_{k_1}^2\omega_n(k)\right).
\eeq%
Identity \eqref{E:mix_der_slave} applies also in the case when the $S_2$
reflection symmetry and the rotational symmetry \eqref{E:material_rot_sym} are
both present for some $\alpha \in (-\pi,-\pi]$, $|\alpha|\neq \pi/2$. Then
\eqref{E:mix_der_slave} holds for $k$ that satisfy
$$
   (k_1,-k_2)=r_{\alpha}(k),
   \text{ i.e.{} } k=\pm |k|e^{-\ri \alpha/2}.
$$

\subsection{Example: Hexagonal Lattice with a Circular Material Structure}%
\label{S:hex_struct}%

\begin{figure}[h!]
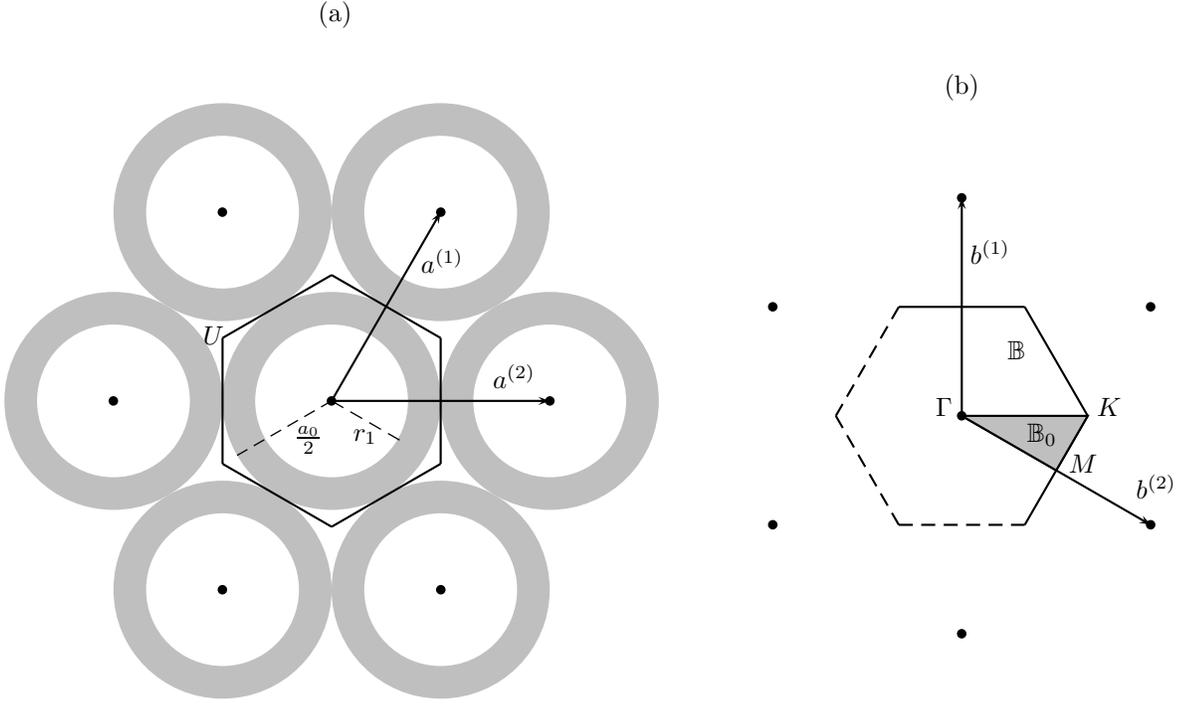

\begin{center}
    \begin{minipage}[c]{0.49\textwidth}
    \centering
    (a)\\
       \psset{unit=2.9cm}
       \pspicture(-1.4,-1.4)(1.7,1.7)
       \pscircle[linecolor=lightgray,fillstyle=solid,fillcolor=lightgray](0,0){0.5}
       \pscircle[linecolor=lightgray,fillstyle=solid,fillcolor=white](0,0){0.36}
       \pscircle[linecolor=lightgray,fillstyle=solid,fillcolor=lightgray](0.5,0.866){0.5}
       \pscircle[linecolor=lightgray,fillstyle=solid,fillcolor=white](0.5,0.866){0.36}
       \pscircle[linecolor=lightgray,fillstyle=solid,fillcolor=lightgray](-0.5,0.866){0.5}
       \pscircle[linecolor=lightgray,fillstyle=solid,fillcolor=white](-0.5,0.866){0.36}
       \pscircle[linecolor=lightgray,fillstyle=solid,fillcolor=lightgray](-0.5,-0.866){0.5}
       \pscircle[linecolor=lightgray,fillstyle=solid,fillcolor=white](-0.5,-0.866){0.36}
       \pscircle[linecolor=lightgray,fillstyle=solid,fillcolor=lightgray](0.5,-0.866){0.5}
       \pscircle[linecolor=lightgray,fillstyle=solid,fillcolor=white](0.5,-0.866){0.36}
       \pscircle[linecolor=lightgray,fillstyle=solid,fillcolor=lightgray](1,0){0.5}
       \pscircle[linecolor=lightgray,fillstyle=solid,fillcolor=white](1,0){0.36}
       \pscircle[linecolor=lightgray,fillstyle=solid,fillcolor=lightgray](-1,0){0.5}
       \pscircle[linecolor=lightgray,fillstyle=solid,fillcolor=white](-1,0){0.36}
       \psline(0.5,0.2887)(0,0.5774)\psline(0,0.5774)(-0.5,0.2887)
       \psline(-0.5,0.2887)(-0.5,-0.2887)\psline(-0.5,-0.2887)(0,-0.5774)
       \psline(0,-0.5774)(0.5,-0.2887)\psline(0.5,-0.2887)(0.5,0.2887)
       \psdots*(0,0)(0.5,0.866)(-0.5,0.866)(-1,0)(-0.5,-0.866)(0.5,-0.866)(1,0)
       \psline{->}(0,0)(1,0)
       \psline{->}(0,0)(0.5,0.866)
       \psline[linewidth=0.5pt,linestyle=dashed](0,0)(-0.433,-0.25)
       \psline[linewidth=0.5pt,linestyle=dashed](0,0)(0.3118,-0.18)
       \put(-0.17,-0.2){$\tfrac{a_0}{2}$}
       \put(0.1,-0.18){$r_1$}
       \put(0.74,0.06){$a^{(2)}$}
       \put(0.41,0.59){$a^{(1)}$}
       \put(-0.59,0.26){$U$}
    \endpspicture
    \end{minipage}
   \begin{minipage}[c]{0.49\textwidth}
    \centering
    (b)\\
    \bigskip
    \bigskip
    \bigskip
       \psset{unit=2.9cm}
       \pspicture(-1,-1)(1,1)
       \psdots*(0,0)(0.866,0.5)(0.866,-0.5)(-0.866,0.5)(-0.866,-0.5)(0,1)(0,-1)
       \put(0.21,0.26){$\B$}
       \psline(0.5774,0)(0.2887,0.5)
       \psline(0.2887,0.5)(-0.2887,0.5)
       \psline[linestyle=dashed](-0.2887,0.5)(-0.5774,0)
       \psline[linestyle=dashed](-0.5774,0)(-0.2887,-0.5)
       \psline[linestyle=dashed](-0.2887,-0.5)(0.2887,-0.5)
       \psline(0.2887,-0.5)(0.5774,0)
       \psline{->}(0,0)(0,1)
       \psline{->}(0,0)(0.866,-0.5)
       \put(0.04,0.7){$b^{(1)}$}
       \put(0.8,-0.38){$b^{(2)}$}
       \put(-0.12,0){$\Gamma$}\put(0.62,0){$K$}\put(0.49,-0.26){$M$}
       \pspolygon[fillstyle=solid,fillcolor=lightgray](0,0)(0.433,-0.25)(0.5774,0)
       \put(0.3,-0.12){$\B_0$}
    \endpspicture
    \end{minipage}
\end{center}
 \caption{\label{F:crystal}\small%
(a) Hexagonal lattice with a cylindrical material structure, (b) the
corresponding first Brillouin zone $\B$ with a shaded irreducible Brillouin
zone $\B_0$. Note that the Brillouin zone has been scaled to fit the figure.}
\end{figure}

As an example we consider the hexagonal lattice in the
$(x_1,x_2)$-plane generated by the vectors
$$
   a^{(1)}= a_0\bspm\cos(\pi/3)\\ \sin(\pi/3)\espm \quad
   \text{and} \quad a^{(2)}
   =a_0\bspm 1\\0\espm \quad \text{with} \quad a_0>0.
$$
In the Wigner--Seitz cell $U$ the material structure is given by the annulus
centered at the lattice point in the origin and having outer and inner radii
$a_0/2$ and $a_0(1.31/4.9)$ respectively. The material properties are given by
$\eta(x) = 2.1025$ for $a_0(1.31/4.9)\leq |x|\leq a_0/2 $ and $\eta(x)=1$
otherwise. This is the same as the crystal used in \cite{AM03}, where the
corresponding band structure was also computed. One choice of vectors
generating the reciprocal lattice is
$$
  b^{(1)}=\tfrac{2\pi}{J_{12}}
  \bspm a^{(2)}_2\\ -a^{(2)}_1\espm= \tfrac{2\pi}{a_0\sin(\pi/3)}
  \bspm 0 \\1\espm, \quad b^{(2)}
  =\tfrac{2\pi}{J_{12}}
  \bspm -a^{(1)}_2\\ a^{(1)}_1\espm= \tfrac{2\pi}{a_0}\bspm 1 \\-\cot(\pi/3)
  \espm,
$$
where $J_{12}=\det(a^{(1)},a^{(2)}) = a^{(1)}_1a^{(2)}_2-a^{(1)}_2a^{(2)}_1$.
These vectors have been obtained via the formulas $\tilde{b}^{(1)}=2\pi
\tfrac{\tilde{a}^{(2)}\times \tilde{a}^{(3)}}{J_{12}}$ and
$\tilde{b}^{(2)}=2\pi \tfrac{\tilde{a}^{(3)}\times \tilde{a}^{(1)}}{J_{12}}$,
where $\tilde{a}^{(j)} = (a^{(j)^T}, 0)^T$, $\tilde{b}^{(j)} = (b^{(j)^T},
0)^T$ for $j\in\{1,2\}$ and $\tilde{a}^{(3)} = (0,0,1)^T$, cf.{}
\cite[Ch.~5]{Ash_Mer_1976}. Figure \ref{F:crystal} shows the crystal geometry
and the corresponding Brillouin zone.

In this case both the rotational symmetry \eqref{E:material_rot_sym} with
$\alpha = \pi/3$, the reflection symmetry \eqref{E:material_refl_sym} as well
as the analogous symmetry $S_2$ do hold. The band structure and Bloch waves can
therefore be recovered via \eqref{E:rot_sym_bands}, \eqref{E:rot_sym_Bloch} and
\eqref{E:refl_sym_bands}, \eqref{E:refl_sym_Bloch} from the irreducible
Brillouin zone $\B_0$ in Figure \ref{F:crystal}, i.e.{} the triangle with
vertices $\Gamma,M,K$, where $\Gamma=(0,0)^T$, $M = \tfrac{1}{2}b^{(2)}$, and
$K=\tfrac{1}{\sqrt{3}}|b^{(2)}|(1,0)^T$. These points are called \emph{high
symmetry points}.

Next we provide some specific information about the values of the second
derivatives of $\omega_n$ at the high symmetry points of the Brillouin zone at
hand using symmetries \eqref{E:der2_rot_sym}, \eqref{E:der2_refl_sym}, and
\eqref{E:der2_refl_symB}. This information will be used in Section
\ref{S:CME_examples}

Identity \eqref{E:der2_rot_sym} with $k=0$ and $\alpha=\pi/3$ yields
\begin{align}\label{E:der2_Gamma}%
  \partial_{k_2}^2\omega_n(\Gamma)=\partial_{k_1}^2\omega_n(\Gamma)
  \quad\text{and }\quad \partial_{k_1,k_2}^2\omega_n(\Gamma)=0
  \qquad \text{for all }n\in \N.
\end{align}%
Symmetry \eqref{E:der2_refl_symB} implies
\beq\label{E:mix_K}%
   \partial_{k_1,k_2}^2\omega_n(K)=0 \qquad \text{for all }n\in \N.
\eeq%
At $k=r_{2\pi/3}(M)$ $(=\tfrac{1}{2}b^{(1)})$ we have $k_1=0$ so that
\eqref{E:der2_refl_sym} implies
\beq\label{E:mix_Mprime}%
  \partial_{k_1,k_2}^2\omega_n(r_{2\pi/3}(M))=0
  \qquad \text{for all }n\in \N.
\eeq%
Relation \eqref{E:mix_der_slave} then yields
\beq\label{E:2nd_der_Mprime}%
  \partial_{k_2}^2\omega_n(r_{2\pi/3}(M))
  = \partial_{k_1}^2\omega_n(r_{2\pi/3}(M))
  \qquad \text{for all }n\in \N.
\eeq%
Applying now \eqref{E:der2_rot_sym} with $\alpha=2\pi/3$, we get
\beq\label{E:M_der}%
  \partial_{k_1}^2\omega_n(M) = \partial_{k_1}^2\omega_n(r_{2\pi/3}(M)),
   \ \partial_{k_2}^2\omega_n(M) = \partial_{k_1}^2\omega_n(r_{2\pi/3}(M)),
   \text{ and } \partial_{k_1,k_2}^2\omega_n(M)=0
\eeq%
for all $n\in\N$. Because $r_{\pi/3}(M)$ is obtained from $r_{2\pi/3}(M)$ by
the reflection $(k_1,k_2)\rightarrow (k_1,-k_2)$, we also have
\begin{align*}
  \partial_{k_1}^2\omega_n(M) = \partial_{k_1}^2\omega_n(r_{\pi/3}(M)),
  \ \partial_{k_2}^2\omega_n(M) = \partial_{k_1}^2\omega_n(r_{\pi/3}(M)),
  \text{ and } \partial_{k_1,k_2}^2\omega_n(r_{\pi/3}(M))=0.
\end{align*}%

As an example we took the configuration from \cite{AM03} as described in
Section \ref{S:hex_struct}. The computations were done with a finite element
Maxwell solver that uses lowest order Nedelec elements \cite{Monk:2003}. These
elements were implemented in the software deal.II \cite{BHK:2007}. The
eigenvalue problems were solved by a Krylov--Schur method.\footnotemark

\footnotetext{SLEPc package (\textsc{http://www.grycap.upv.es/slepc/})}

We computed the eigenvalues $\{\omega_n(k)\}_{n=1,14}$ and corresponding
eigenfunctions $\{p_n(k,\cdot)\}_{n=1,14}$ for each vertex $k$ in a
discretization of the Brillouin zone $\B$. The error level of this computations
is about $10^{-3}$ in the curl-norm and it is estimated from a series of
computations on a sequence of nested grids.

\begin{figure}[h!]
\begin{center}
 \epsfig{figure = 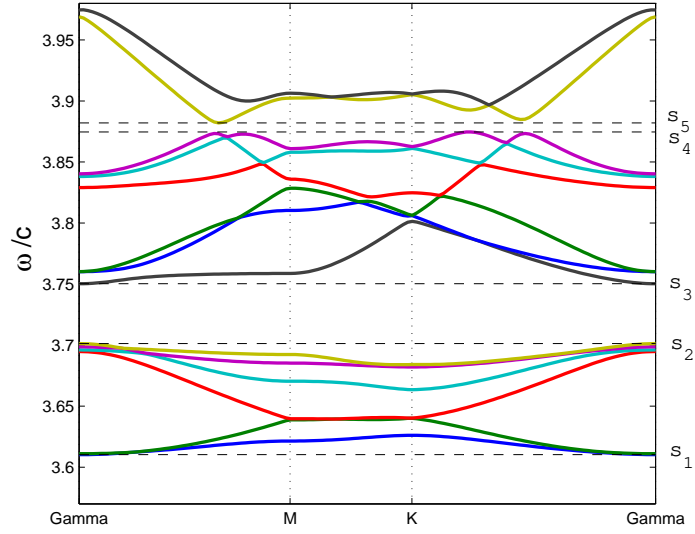,scale=0.62}
\caption{\label{F:band_str} \small Band structure $k\mapsto\omega_n(k)$
for the described hexagonal lattice
with the cylindrical material structure: the first 14 eigenvalues along
$\partial\B_0$. Three band gaps appear on the positive half axis $\Omega\geq 0$:
one between $0$ and $\omega_1$, one between $\omega_6$ and $\omega_7$, and one
between $\omega_{12}$ and $\omega_{13}$. Gap edges are marked by
$s_1, \ldots, s_5$.}
\end{center}
\end{figure}

In Figure \ref{F:band_str} we present the numerically computed band structure
over $\partial\B_0$ (following the tradition) for the above described crystal
and for $\kappa=5(2\pi/a_0)$. Here, $\partial\B_0$ is represented by 128
$k$-points. It has, however, been checked that the observed gaps do not get
narrower in the interior of $\B$. Three band gaps appear on the positive half
of the $\omega$ axis, one between $0$ and $\omega_1$, another one between
$\omega_6$ and $\omega_7$ and the last one between $\omega_{12}$ and
$\omega_{13}$.

To get the extremal points at the band edges we used a bisection method in $k$
which was initialised with data obtained from the band structure computation.
The approximations to 1st and 2nd order derivatives of $k\mapsto\omega_{n}(k)$
at the extremal values were obtained by first projecting
$k\mapsto\omega_{n}(k)$ onto a locally quadratic finite element space and then
taking mean values of the derivatives around vertices.

\section{Derivation of Coupled Mode Equations for Gap Solitons near Band
Edges}\label{S:CME_deriv}

\subsection{Slowly varying envelope approach}%
\label{S:slowly envelope}%

We seek gap solitons $E$ of \eqref{E:NL_Maxw}. Afterward, the full electric
field can be recovered via \eqref{E:field_form}.

In the following let us assume that
\begin{itemize}
   \item[(A1)]
the spectrum $\{\omega_n(k):k\in\B,\,n\in\N\}$ possesses a gap,
   \item[(A2)]%
one of the gap edges, denoted by $\omega_*$, is attained at precisely $N\in\N$ points
$k^{(1)},\ldots,k^{(N)}\in \B$ by bands with indices
$n_1,\ldots,n_N$, respectively, where the $k$-points and/or band indices
are not necessarily all distinct,
   \item[(A3)]%
for each $j\in \{1, \ldots, N\}$ the band $\omega_{n_j}$ is twice
continuously differentiable in $k$ at $k=k^{(j)}$,
   \item[(A4)]%
$\pa_k^2\omega_{n_j}(k^{(j)})$, the Hessian of $\omega_{n_j}$ at
$k=k^{(j)}$, is (positive or negative) definite for each $j\in \{1, \ldots,
N\}$.
\end{itemize}
The smoothness assumption (A3) is needed to justify our Taylor expansions of
$\omega_{n_j}$ near $k^{(j)}$. Bands $\omega_n$ are generally only Lipschitz
continuous due to possible transversal intersections of bands and their
numbering according to size \cite{MP_1996}.
Away from points of intersection or tangency bands can be shown to be actually
analytic in $k$ by standard perturbation theory \cite{Kato_1995}. The simplest
situation when (A3) is satisfied is thus when each band $\omega_{n_j}$ is
isolated near $k^{(j)}$, which is equivalent to $n_1=\ldots=n_N$ due to our
ordering of bands according to size of $\omega_n(k)$ at each $k$.

Note that since each band $\omega_{n_j}$ has an extremum at $k=k^{(j)}$, we
have $\pa_{k_1}\omega_{n_j}(k^{(j)})=\pa_{k_2}\omega_{n_j}(k^{(j)})=0$ for
$j\in \{1,\ldots,N\}$. Assumption (A4) then guarantees that the leading order
terms in the Taylor expansion of the band $\omega_{n_j}$ around $k=k^{(j)}$ are
in fact quadratic.

The asymptotic expansion for the electrical field $E$ of gap solitons with
$\omega$ in the gap and in the vicinity of the edge $\omega_*$ is expected
\cite{DPS09,DU09} to be of the following \emph{slowly varying envelope} form
\begin{gather}\label{E:ansatz_phys}
\begin{split}
  &\eps\sum_{j=1}^NA_j(y)u_{n_j}(k^{(j)};x)
   +\eps^2\psi^{(1)}(x)
   +\eps^3\psi^{(2)}(x)+\mathcal{O}(\eps^4),\\
  &\omega = \omega_*+\Omega\eps^2, \quad y
   = \eps x, \quad 0<\eps \ll 1,
\end{split}
\end{gather}
where $A_j:\R^2\to\C$ is a fast decaying smooth function and where $\Omega =
\pm 1$. The sign of $\Omega$ is determined by the condition that
$\omega_*+\eps^2 \Omega$ lies in the gap.

Performing a multiple scales analysis in the physical variables $(x,y)$ is
impossible. The reason is that in order to solve the resulting equations at
each order of the expansion, one has to ensure that inhomogeneous terms are
orthogonal to the kernel of $L-\omega_*^2\eta$, i.e., to
$u_{n_j}(k^{(j)};\pkt)$ for all $j \in \{1, \ldots, N\}$. This orthogonality
needs to be checked on the common period of those $u_{n_j}$. If, however, one
of the components of $k^{(j)}$ is irrational, the corresponding $u_{n_j}$ is
not even periodic and this approach fails similarly to \cite{DU09}. We
therefore perform the asymptotic analysis in Bloch variables where all
functions are $U$-periodic in $x$ and orthogonality conditions are always posed
over $U$.

Let us define the \emph{Bloch transform} $\CT:E\mapsto\wt{E}$ and its inverse,
cf.{} \cite[Ch.~7]{BCM_2001}, by
\begin{gather}
\begin{split}\notag
  \wt{E}(k;x) &= (\CT E)(k;x)
   = \sum_{K\in\Lam^*} e^{\ri K\cdot x}\wh{E}(k+K),
     \quad E(x) = (\CT^{-1} \wt{E})(x)
   = \int_{\B}e^{\ri k\cdot x} \wt{E}(k;x)\dd k
\end{split}
\end{gather}
for all $x,k\in\R^2$, where $\wh{E}$ denotes the \emph{Fourier transform\/} of
$E$
\begin{gather}\label{E:FT}
 \notag
 \wh{E}(k):= (\CF E)(k)
 :=\frac{1}{(2\pi)^2}\int_{\R^2}E(x)e^{-\ri k\cdot x}\dd x,
 \quad
 E(x)=(\CF^{-1}\wh{E})(k)
 :=\int_{\R^2}\wh{E}(k)e^{\ri k\cdot x}\dd k.
\end{gather}
By definition we have the following properties of the Bloch transform
\begin{alignat}{2}
  \notag
  \wt{E}(k;x+R)
  &= \wt{E}(k;x) &&\qquad\text{for all }R \in \Lam, \\
  \label{E:Bloch_B_per_in_k}
  \wt{E}(k+K;x)
  &= e^{-\ri K\cdot x}\wt{E}(k;x) &&\qquad\text{for all }K \in \Lam^*.
\end{alignat}
Multiplication of two functions $f,g$ in physical space corresponds to
convolution in Bloch space, i.e.,
\begin{gather}
  \notag
  \big(\CT (fg)\big)(k;x)
   = \int_{\B}\wt{f}(k-l;x)\wt{g}(l;x)\dd l
   =: \big(\wt{f} *_{\Bsm}\wt{g}\big)(k;x),
\end{gather}
where \eqref{E:Bloch_B_per_in_k} is used if $k-l\notin\B$. Especially, if
$x\mapsto f(x)$ is $U$-periodic, then
\[
  \big(\CT (fg)\big)(k;x) = f(x)(\CT g)(k;x).
\]%
This can be easily checked by writing $f$ in the form of a Fourier series,
i.e.{} $f(x) = \sum_{K\in\Lam^*}c_K e^{\ri K\cdot x}$, cf.{}
\cite[Ch.~7]{BCM_2001}. Exploiting this observation and applying the Bloch
transform to \eqref{E:NL_Maxw} leads to
\beq
 \notag
 \left(\wt{L}(k)-\omega^2\eta(x)\right)\wt{E}(k;x)
 = \omega^2 \wt{P}_\text{NL}(k;x)\quad\text{for all }x,k\in\R^2,
\eeq%
where
\begin{align*}
   \wt{P}_\text{NL}(k;\pkt)
   = \chici\CT\big(2|E|^2\,E+E\cdot E \,\barl{E}\big)
   = \chici\big(2(E\,.\!*_\Bsm\barl{E})*_\Bsm E
      +(E\,.\!*_\Bsm E)*_\Bsm\barl{E}\big),
\end{align*}
with $f\,.\!*_\Bsm g:=\sum_jf_j*_\Bsm g_j$ for vector valued $f,g$, while
$f*_\Bsm g$ is understood componentwise for scalar $f$ and vector valued $g$.
By definition of the Bloch-- and Fourier transformation one immediately finds
\[
   \CT\big(A_j(\eps\pkt)e^{\ri k^{(j)}\cdot(\pkt)}\big)(k;x)
   = \eps^{-2}\sum_{K\in\Lam^*}\wh{A}_j
   \left(\tfrac1{\eps}(k-k^{(j)}+K)\right)e^{\ri K\cdot x},
\]%
so that the asymptotic ansatz \eqref{E:ansatz_phys} is transformed to
\beq\label{E:ansatz_Bloch}%
  \eps\wt{\psi}^{(0)}(k;x)
  +\eps^2\wt{\psi}^{(1)}(k;x)+\eps^3\wt{\psi}^{(2)}(k;x)
  +O(\eps^4),
\eeq%
where
$$
 \wt{\psi}^{(0)}(k;x)=\eps^{-2}\sum_{j=1}^N
 \sum_{K \in \Lam^*}\wh{A}_j\left(\tfrac1{\eps}(k-k^{(j)}+K)\right)
 e^{\ri K \cdot x}p_{n_j}(k^{(j)};x).
$$
Similarly to \cite{DU09} and \cite{DU_err11}, due to the fast decay of the
Bloch transform of $A_j$ in $k$, we approximate
$\wh{A}_j\left(\tfrac1{\eps}(k-k^{(j)}+K)\right)$ by
$\chi_{D_{\eps^r}}^{}\left(k-k^{(j)}+K\right)
\wh{A}_j\left(\tfrac1{\eps}(k-k^{(j)}+K)\right)$ for some $r\in(0,1)$, where
$\chi_{S}$ is the indicator function of a set $S$, $D_\delta:=B_\delta(0)$ with
$B_\delta(z) :=\{k\in\R^2:|k-z|<\delta\}$ for $\delta>0$, $z\in\R^2$.

We will therefore introduce the approximation
\[
   \wt{E}(k;x) = \eps^{-1}\wt{E}^{(0)}(k;x)
     + \wt{E}^{(1)}(k;x)+\eps \wt{E}^{(2)}(k;x) + O(\eps^2)
\]%
with
\beq 
   \notag
   \wt{E}^{(0)}(k;x)
   =\sum_{j=1}^N\sum_{K\in\Lam^*}
    \chi_{D_{\eps^r}}^{}\left(k-k^{(j)}+K\right)
    \wh{A}_j\left(\tfrac1{\eps}(k-k^{(j)}+K)\right)e^{\ri K\cdot x}
    p_{n_j}(k^{(j)};x)
\eeq%
for all $k\in\B$ and $x\in\R^2$. In the following we will use the notation
$K^m=m_1b^{(1)}+m_2b^{(2)}\in\Lam^*$ for $m\in\Z^2$ for convenience. As an
abbreviation we let $\ell^{(j,m)}(k):=\tfrac1{\eps}(k-k^{(j)}+K^m)$ for
$k\in\R^2$ and $m\in\Z^2$, so that $\wt{E}^{(0)}$ is given as
\beq \label{E:E0_approx}%
   \wt{E}^{(0)}(k;x)
   =\sum_{j=1}^N\sum_{m\in\Z^2}
    \chi_{D_{\eps^r}}^{}\big(\ell^{(j,m)}(k)\big)
    \wh{A}_j\big(\ell^{(j,m)}(k)\big)e^{\ri K^m\cdot x}
    p_{n_j}(k^{(j)};x).
\eeq%
Note that $\wt{E}^{(0)}(\pkt;x)$ is supported on a set of (for sufficiently
small $\eps$) disjoint balls $B_{\eps^r}(k^{(j)}-K^m)$, $j\in\{1,\ldots,N\}$,
$m\in\Z^2$.

\subsection{Formal asymptotic analysis}%
\label{S:asymptotic_analysis}

Let us proceed with a formal asymptotic analysis of \eqref{E:NL_Maxw}. First,
we consider $k$ close to $k^{(j)}-K^m$, i.e., $k\in B_{\eps^r}(k^{(j)}-K^m)$
for some $j\in \{1, \ldots, N\}, m\in \Z^2$. Then
\beq\label{E:Ltil_expand}%
\begin{split}
   \wt{L}(k)
   &= \wt{L}\big(k^{(j)}-K^m+\eps\ell^{(j,m)}(k)\big)\\
   &= \wt{L}(k^{(j)}-K^m)
     +\eps \ell^{(j,m)}(k)\cdot\pa_{k}\wt{L}(k^{(j)}-K^m)
     +\frac12\eps^2Q(\ell^{(j,m)}(k)),
\end{split}
\eeq%
where we have used the fact that the second derivatives of $\wt{L}$ are
constant in $k$, see \eqref{E:Ltil_2nd_der}, and where
\[
\begin{split}
  \ell^{(j,m)}(k)\cdot\pa_{k}\wt{L}(k^{(j)}-K^m)
  &= \sum_{i=1}^2\ell^{(j,m)}_i(k)\pa_{k_i}\wt{L}(k^{(j)}-K^m), \text{ and}\\
  Q(\ell^{(j,m)}(k))
  & = \sum_{a,b=1}^2\ell_a^{(j,m)}(k)\ell_b^{(j,m)}(k)\pa^2_{k_a,k_b}\tilde{L}.
\end{split}
\]

Using \eqref{E:ansatz_Bloch}, \eqref{E:E0_approx}, \eqref{E:Ltil_expand} and
$\omega = \omega_* + \Omega \eps^2$, we get a hierarchy of equations at each
power of $\eps$ for $x\in U$ and $k\in B_{\eps^r}(k^{(j)}-K^m)$. We now study
the equations related to $\eps^{-1},\eps^0,\eps^1$ under the condition that the
nonlinear term contributes to $\eps^1$, which is confirmed later in
\eqref{E:NL_term_conv}.

$\mbf{O(\eps^{-1})}$: The resulting equation is
\begin{align*}
   &\wh{A}_j\big(\ell^{(j,m)}(k)\big)
    \left(\wt{L}(k^{(j)}-K^m)-\omega_*^2\eta(x)\right)
    \big(p_{n_j}(k^{(j)};x)e^{\ri K^m\cdot x}\big)\\
   &\qquad=\wh{A}_j\big(\ell^{(j,m)}(k)\big)e^{\ri K^m\cdot x}
    \left(\wt{L}(k^{(j)})-\omega_*^2\eta(x)\right)
    p_{n_j}(k^{(j)};x)\solleq 0.
\end{align*}%
This holds by the definitions of $\omega_*=\omega_{n_j}(k^{(j)})$ and
$p_{n_j}(k^{(j)};\pkt)$.

$\mbf{O(1)}$: The resulting equation is
\begin{align*}
   &\left(\wt{L}(k^{(j)}-K^m)-\omega_*^2\eta(x)\right)\wt{E}^{(1)}(k;x)\\
   &\qquad=-\wh{A}_j(\ell^{(j,m)}(k)\big)\big(\ell^{(j,m)}(k)\cdot
    \pa_{k}\wt{L}(k^{(j)}-K^m)\big)\big(p_{n_j}(k^{(j)};x)e^{\ri K^m\cdot x}\big)\\
   &\qquad=-\wh{A}_j\big(\ell^{(j,m)}(k)\big)e^{\ri K^m\cdot x}
    \big(\ell^{(j,m)}(k)\cdot\pa_{k}\wt{L}(k^{(j)})\big)p_{n_j}(k^{(j)};x)
    \solleq 0.
\end{align*}%
Using \eqref{E:deriv_eq}, the solution is found to be
\beq\label{E:E1}%
   \wt{E}^{(1)}(k;x)
   =\wh{A}_j(\ell^{(j,m)}(k))e^{\ri K^m\cdot x}
    \big(\ell^{(j,m)}(k)\cdot\pa_{k}p_{n_j}(k^{(j)};x)\big),
\eeq%
where $\ell^{(j,m)}(k)\cdot\pa_{k}p_{n_j}(k^{(j)};x) =
\sum_{i=1}^2\ell^{(j,m)}_i(k)\pa_{k_i}p_{n_j}(k^{(j)};x)$.

$\mbf{O(\eps)}$: The contribution of $\wt{L}(k)\wt{E}$ is
\begin{align*}
   &\left(\wt{L}(k^{(j)}-K^m)-\omega_*^2\eta(x)\right)\wt{E}^{(2)}(k;x)\\
   &\qquad{}+\tfrac{1}{2}Q(\ell^{(j,m)})\wt{E}^{(0)}(k;x)
    -2\omega_*\Omega\eta(x)\wt{E}^{(0)}(k;x)\\
   &\qquad{}+\big(\ell^{(j,m)}(k)\cdot\pa_{k}\wt{L}(k^{(j)}-K^m)\big)
    \wt{E}^{(1)}(k;x).
\end{align*}%
By insertion of the previous results this gives (for $k\in
B_{\eps^r}(k^{(j)}-K^m)$)
\begin{align}
   \notag
   &\left(\wt{L}(k^{(j)}-K^m)-\omega_*^2\eta(x)\right)\wt{E}^{(2)}(k;x)\\
   \notag
   &\qquad\quad{}+\Big[\tfrac{1}{2}Q(\ell^{(j,m)}(k))p_{n_j}(k^{(j)};x)
     -2\omega_*\Omega\eta(x)p_{n_j}(k^{(j)};x)\\
   \notag
   &\qquad\qquad\quad{}+\big(\ell^{(j,m)}(k)\cdot\pa_{k}\wt{L}(k^{(j)}-K^m)\big)
    \big(\ell^{(j,m)}(k)\cdot\pa_{k}p_{n_j}(k^{(j)};x)\big)\Big]\wh{A}_j
    \big(\ell^{(j,m)}(k)\big)e^{\ri K^m\cdot x}\\
   \label{E:Oeps}
   &\qquad\solleq\omega_*^2\chici(x)\frac1{\eps^4}
    \Big(2\big(\wt{E}^{(0)}\,.\!*_\Bsm\wt{\barl{E^{(0)}}}\big)*_\Bsm \wt{E}^{(0)}
    +\big(\wt{E}^{(0)}\,.\!*_\Bsm \wt{E}^{(0)}\big)*_\Bsm\wt{\barl{E^{(0)}}}\Big)
    (k;x)\\
   \notag
   &\qquad=:\omega_*^2\chici(x)\wt{G}_j(k,x).
\end{align}%

The remainder of the section is devoted to the analysis of the structure of
$\wt{G}_j$ in \eqref{E:Oeps} and to the derivation of a solvability condition
for \eqref{E:Oeps}.

Let us first analyze the nonlinearity. The convolutions in \eqref{E:Oeps} can
be expanded into the form
\beq\label{E:NL_termE}%
  \wt{E}^{(0)}_a *_{\Bsm} \wt{E}^{(0)}_b *_{\Bsm} \wt{\barl{E^{(0)}_c}}
  = \sum_{\alpha,\beta,\gamma=1}^N \xi_{\alpha,a} *_{\Bsm}
    \xi_{\beta,b} *_{\Bsm} \xi_{\gamma,c}^\times,
\eeq%
where $a,b,c \in \{1,2,3\}$, and functions $\xi_{\alpha,a}$ and
$\xi_{\alpha,a}^\times$ are given by
\[%
\begin{array}{ll}
 & \xi_{\alpha,a}(k;x)
   := p_{n_\alpha,a}(k^{(\alpha)};x) \sum\limits_{z\in \Z^2}
   \chi_{D_{\eps^r}}^{}\big(k-k^{(\alpha)}+K^{z}\big)\wh{A}_\alpha
   \left(\tfrac1{\eps}(k-k^{(\alpha)}+K^{z})\right)e^{\ri K^{z}\cdot x},
   \\[6pt]
 & \xi_{\alpha,a}^\times(k;x)
   := \barl{p_{n_\alpha,a}}(k^{(\alpha)};x) \sum\limits_{z\in\Z^2}
   \chi_{D_{\eps^r}}^{}\big(k+k^{(\alpha)}-K^{z}\big)\wh{A}_\alpha
   \left(\tfrac1{\eps}(k+k^{(\alpha)}-K^{z})\right)e^{-\ri K^{z}\cdot x}.
\end{array}
\]%
Note that \eqref{E:NL_termE} represents all the nonlinear terms in
\eqref{E:Oeps} due to commutativity of $*_{\Bsm}$. The summands in
\eqref{E:NL_termE} have the form
\beq\label{E:double_convol}%
   (\xi_{\alpha,a}*_{\Bsm}\xi_{\beta,b}*_{\Bsm}\xi_{\gamma,c}^\times)(k;x)
   = \sum_{n,o,q\in\Z^2}g_{noq}(k;x)
   = \sum_{n\in M_\alpha^{(2)},\,o\in M_\beta^{(2)},\,q\in M_\gamma}
     g_{noq}(k;x),
\eeq%
where (with indices $\alpha,\beta,\gamma,a,b,c$ suppressed)
\beq\label{E:NL_term}%
\begin{split}
 g_{noq}(k;x)
 &= e^{\ri(K^n+K^o-K^q)\cdot x} p_{n_{\alpha},a}(k^{(\alpha)};x)
  p_{n_{\beta},b}(k^{(\beta)};x)\ov{p_{n_{\gamma},c}}(k^{(\gamma)};x)\\
 &\quad \int\limits_{\B}\int\limits_{\B}
  \chi_{D_{\eps^r}}^{}\big(k-t-k^{(\alpha)}+K^n\big)
  \wh{A}_\alpha\left(\tfrac1{\eps}(k-t-k^{(\alpha)}+K^n)\right)\\
 &\qquad\qquad \times
  \chi_{D_{\eps^r}}^{}\big(t-s-k^{(\beta)}+K^o\big)
  \wh{A}_\beta\left(\tfrac1{\eps}(t-s-k^{(\beta)}+K^o)\right)\\[4pt]
 &\qquad\qquad \times
  \chi_{D_{\eps^r}}^{}\big(s+k^{(\gamma)}-K^q\big)
  \wh{\barl{A}}_\gamma\left(\tfrac1{\eps}(s+k^{(\gamma)}-K^q)\right)
  \dd s\dd t
\end{split}
\eeq%
and with
\begin{align*}%
   M_\gamma &= \{z\in\Z^2 : k-k^{(\gamma)}+K^{z}\in B_{\eps^r}(0)
      \text{ for some } k\in \B \text{ and all } \eps>0 \},\\
   M_\flat^{(2)} &= \{z\in\Z^2 : k-k^{(\flat)}+K^{z}\in B_{\eps^r}(0)
      \text{ for some } k\in \B + \B \text{ and all } \eps>0 \}
\end{align*}%
for $\flat\in\{\alpha,\beta\}$. The truncation of the series in
\eqref{E:double_convol} comes from the fact that for $s,t,k \in \B$ we have
$t-s\in\B + \B$ and $k-t \in \B + \B$ so that the three characteristic
functions in \eqref{E:NL_term} can be nonzero only for $n\in M_\alpha^{(2)}$,
$o\in M_\beta^{(2)}$, and $q\in M_\gamma$. More precisely, this is seen as
follows.

Only those combinations of $n,o,q$ which produce nonzero values of all the
three characteristic functions in \eqref{E:NL_term} and of the function
$\chi_{D_{\eps^r}}^{}\big(\pkt-k^{(j)}+K^m\big)$ in \eqref{E:Oeps} for given
$j$ and some $k,t,s\in \B$ are of relevance. Firstly,
$\chi_{D_{\eps^r}}^{}\big(s+k^{(\gamma)}-K^q\big)$ is nonzero for some $s\in
\B$ and for arbitrary $\eps>0$ if and only if $s_0:=-k^{(\gamma)}+K^q\in
\overline{\B}$ (the closure of $\B$) for some $q\in\Z^2$, which is equivalent
to
\beq\label{E:q_cond}%
   q\in M_\gamma.
\eeq%
Secondly, for a fixed $q$ the factor
$\chi_{D_{\eps^r}}^{}\big(t-s-k^{(\beta)}+K^o\big)$ is nonzero for all $\eps
>0$ and some $t\in\B$ and $s$ obtained in the first step if and only if
$t_0:=s_0+k^{(\beta)}-K^o \in \overline{\B}$, i.e.,
\beq\label{E:o_cond}%
   k^{(\beta)}-k^{(\gamma)}+K^q-K^o \in \barl{\B}.
\eeq%
This can always be satisfied by a choice of $o\in M^{(2)}_\beta$. Finally, for
fixed $q$ and $o$ we need that
$\chi_{D_{\eps^r}}^{}\big(k-t-k^{(\alpha)}+K^n\big)$ does not vanish for some
$k\in\B$ with $k-k^{(j)}+K^m\in D_{\eps^r}$ and all $\eps>0$, where this latter
restriction is due to the restriction $k\in B_{\eps^r}(k^{(j)}-K^m)$ in
\eqref{E:Oeps}. In other words, we need that $k_0:=k^{(j)}-K^m\in\barl{\B}$ and
$0=k_0-t_0-k^{(\alpha)}+K^n$, i.e.{}
\beq\label{E:n_cond}%
   k^{(\alpha)}+k^{(\beta)}-k^{(\gamma)}+K^q-K^o-K^n = k^{(j)}-K^m
      \in\barl{\B}
\eeq%
for some $n\in \Z^2$. In fact, all solutions for $n$ of \eqref{E:n_cond} lie
in $M_\alpha^{(2)}$.

In summary, for $\alpha,\beta, \gamma \in \{1, \ldots, N\}$ the term $g_{noq}$
is nonzero in \eqref{E:NL_term} if $n,o,q$ satisfy \eqref{E:q_cond},
\eqref{E:o_cond} and \eqref{E:n_cond}. So the term $\xi_{\alpha,a} *_{\Bsm}
\xi_{\beta,b} *_{\Bsm} \xi_{\gamma,c}^\times$ enters $\wt{G}_j$ provided
\[%
   \CA_{\alpha,\beta,\gamma,j}
      :=\Big\{(n,o,q)\in (\Z^2)^3 : n\in M_\alpha^{(2)},\,o\in M_\beta^{(2)},
      \,q\in M_\gamma\text{ and } \eqref{E:o_cond},\,\eqref{E:n_cond}
      \text{ hold}\Big\}
\]%
is nonempty for some $m\in M_j$. Note that we omitted an index $m$ in this
definition, because the set is either nonempty or empty for all $m\in M_j$.
Indeed, if $m$ is one index that meets the requirements with $(n,o,q)$ and $z$
is any other index in $M_j$, then $z$ meets the requirements for $(n+m-z,o,q)$.
$\CA_{\alpha,\beta,\gamma,j}$ can be constructed by a computer code that scans
all possible combinations of $n,o,q$. This will be discussed in Section
\ref{S:CME_examples}.

Due to the characteristic function, the integration domains in
\eqref{E:NL_term} can be reduced to $s\in B_{\eps^r}(-k^{(\gamma)}+K^q)\cap \B$
and $t\in B_{2\eps^r}(k^{(\beta)}-k^{(\gamma)}-K^o+K^q) \cap \B$. Now we
introduce the change of variables $\tilde{s} :=(s+k^{(\gamma)}-K^q)/\eps$ and
$\tilde{t} := (t-k^{(\beta)}+k^{(\gamma)}+K^o-K^q)/\eps$ to get
\beq\label{E:NL_term_conv}%
\begin{split}
   &g_{noq}(k;x)
   =\eps^4 e^{\ri(K^n+K^o-K^q)\cdot x}p_{n_{\alpha},a}(k^{(\alpha)};x)
    p_{n_{\beta},b}(k^{(\beta)};x)\ov{p_{n_{\gamma},c}}(k^{(\gamma)};x)\\
   &\qquad\times\int
    \limits_{D_{2\eps^{r-1}}\cap\tfrac{\B-k^{(\beta)}+k^{(\gamma)}+K^o-K^q}{\eps}}
    \int\limits_{D_{\eps^{r-1}}\cap\tfrac{\B+k^{(\gamma)}-K^q}{\eps}}
    \chi_{D_{\eps^{r-1}}}^{}\left(
    \tfrac{k-(k^{(\alpha)}+k^{(\beta)}-k^{(\gamma)})+K^n+K^o-K^q}{\eps}
    -\tilde{t}\right)\\
   &\qquad\quad\quad\times\wh{A}_\alpha\left(
    \tfrac{k-(k^{(\alpha)}+k^{(\beta)}-k^{(\gamma)})+K^n+K^o-K^q}{\eps}
    -\tilde{t}\right)
    \chi_{D_{\eps^{r-1}}}^{}(\tilde{t}-\tilde{s})\wh{A}_\beta(\tilde{t}-\tilde{s})
    \chi_{D_{\eps^{r-1}}}^{}(\tilde{s})\wh{\barl{A}}_\gamma(\tilde{s})
    \dd\tilde{s} \dd\tilde{t}.
\end{split}
\eeq%
The factor $\eps^4$ in this formula shows that $\wt{G}_j=O(1)$ as required for
the consistent asymptotic expansion. If \eqref{E:n_cond} is satisfied,
\eqref{E:NL_term_conv} becomes
\beq\label{E:NL_term_conv2}%
\begin{split}
   &g_{noq}(k;x)
   =\eps^4 e^{\ri(k^{(\alpha)}+k^{(\beta)}-k^{(\gamma)}-k^{(j)}+K^m)\cdot x}
    p_{n_{\alpha},a}(k^{(\alpha)};x)
    p_{n_{\beta},b}(k^{(\beta)};x)\ov{p_{n_{\gamma},c}}(k^{(\gamma)};x) \\
   &\qquad\times\int
    \limits_{D_{2\eps^{r-1}}\cap\frac{\B-k^{(\beta)}+k^{(\gamma)}+K^o-K^q}\eps}
    \int\limits_{D_{\eps^{r-1}}\cap\frac{\B+k^{(\gamma)}-K^q}\eps}
    \chi_{D_{\eps^{r-1}}}^{}\left(\tfrac{k-k^{(j)}+K^m}{\eps}-\tilde{t}\right)\\
   &\qquad\qquad\times\wh{A}_\alpha\left(\tfrac{k-k^{(j)}+K^m}{\eps}-\tilde{t}\right)
    \chi_{D_{\eps^{r-1}}}^{}(\tilde{t}-\tilde{s})\wh{A}_\beta(\tilde{t}-\tilde{s})
    \chi_{D_{\eps^{r-1}}}^{}(\tilde{s})\wh{\barl{A}}_\gamma(\tilde{s})
    \dd\tilde{s} \dd\tilde{t}
\end{split}
\eeq%
for $k\in B_{\eps^r}(k^{(j)}-K^m)$. As we show in Remark \ref{R:convol_sum},
summing, for fixed $k,j,m$, the terms \eqref{E:NL_term_conv2} over $(n,o,q)\in
\CA_{\alpha,\beta,\gamma,j}$ yields a double convolution integral in
$\tilde{s},\tilde{t}$ over the full discs $D_{2\eps^{r-1}}$ and
$D_{\eps^{r-1}}$, i.e.,
\beq\label{E:NL_term_conv_sum}%
\begin{split}
   &\big(\xi_{\alpha,a} *_{\Bsm} \xi_{\beta,b} *_{\Bsm}
    \xi_{\gamma,c}^\times\big)(k;x)
    =\eps^4 e^{\ri(k^{(\alpha)}+k^{(\beta)}-k^{(\gamma)}-k^{(j)}+K^m)\cdot x}
    p_{n_{\alpha},a}(k^{(\alpha)};x)p_{n_{\beta},b}(k^{(\beta)};x)
    \barl{p_{n_{\gamma},c}}(k^{(\gamma)};x) \\
   &\quad\times\int
    \limits_{D_{2\eps^{r-1}}}\int\limits_{D_{\eps^{r-1}}}
    \chi_{D_{\eps^{r-1}}}^{}\big(\ell^{(j,m)}(k)-\tilde{t}\big)
    \wh{A}_\alpha\big(\ell^{(j,m)}(k)-\tilde{t}\big)
    \chi_{D_{\eps^{r-1}}}^{}(\tilde{t}-\tilde{s})
    \wh{A}_\beta(\tilde{t}-\tilde{s})
    \chi_{D_{\eps^{r-1}}}^{}(\tilde{s})\wh{\barl{A}}_\gamma(\tilde{s})
    \dd\tilde{s}\dd\tilde{t}\\
   &\quad=:\eps^4 e^{\ri(-k^{(j)}+K^m)\cdot x}
    u_{n_\alpha,a}(k^{(\alpha)};x)u_{n_\beta,b}(k^{(\beta)};x)
    \barl{u_{n_\gamma,c}}(k^{(\gamma)};x)\;
    \tilde{h}_{\alpha,\beta,\gamma}^{(\eps)}(\ell^{(j,m)}(k))
\end{split}
\eeq%
for $k\in B_{\eps^r}(k^{(j)}-K^m)$. Here we have used
$u_{n_{\alpha},a}(k^{(\alpha)};x) =p_{n_{\alpha},a}(k^{(\alpha)};x)e^{\ri
k^{(\alpha)}\cdot x}$, etc., and we defined
$\tilde{h}_{\alpha,\beta,\gamma}^{(\eps)}(\ell^{(j,m)}(k))$ as an abbreviation
for the integral on the right hand side.

\brem\label{R:convol_sum} To show that the sum of $g_{noq}$ over $(n,o,q)\in
\CA_{\alpha,\beta,\gamma,j}$ yields a double convolution integral over full
discs, let us first note that the definitions of $M_\gamma$ and $M^{(2)}_\beta$
ensure
\beq\label{E:Mgamma_sum}
  \bigcup_{q\in M_\gamma}
  \left((\B+k^{(\gamma)}-K^q)\cap D_{\eps^r}\right)=D_{\eps^r},
\eeq%
and
\beq\label{E:M2beta_sum}
  \bigcup_{o\in M^{(2)}_\beta}
  \left((\B-k^{(\beta)}+k^{(\gamma)}+K^o-K^q)\cap D_{2\eps^r}\right)
  =D_{2\eps^r}.
\eeq%
These are obvious when $k^{(\gamma)} \in \operatorname{int}(\B)$ and
$k^{(\beta)},k^{(\gamma)} \in \operatorname{int}(\B)$, respectively, because
then $M_\gamma=M^{(2)}_\beta=\{(0,0)^T\}$. But when $k^{(\gamma)}\in\pa\B$,
then only a fraction of $-k^{(\gamma)}+D_{\eps^r}$ lies in $\B$ (in our example
with a hexagonal $\B$ the fraction is a half unless $k^{(\gamma)}$ is a vertex
of $\B$, in which case it is a third) and the rest lies in periodicity cells
centered at neighboring reciprocal lattice points. Each point $\ell$ in this
rest is therefore mapped to $\B$ via $\ell+K^q$ with some $q\in M_\gamma$, and
we thus have \eqref{E:Mgamma_sum}. By an analogous argument, observing that
$k^{(\beta)}-(k^{(\gamma)}-K^q)\in\B +D_{\eps^r}$ for all $q\in M_\gamma$, we
get \eqref{E:M2beta_sum} from the definition of $M^{(2)}_\beta$.

Let us now assume \eqref{E:n_cond} and show that for each $K^q$ fixed, i.e.{}
for each fixed integration domain in the inner integral in
\eqref{E:NL_term_conv}, the sum of $g_{noq}$ over $(n,o,q)\in
\CA_{\alpha,\beta,\gamma,j}$ yields an integration over the full disc
$D_{2\eps^r}$ in the outer integral. If this were not the case, i.e.{} if
$\exists\ell\in D_{2\eps^r}$ such that $\ell\notin
\B-k^{(\beta)}+k^{(\gamma)}+K^o-K^q$ for any such $(n,o,q)\in
\CA_{\alpha,\beta,\gamma,j}$, then by \eqref{E:M2beta_sum} there would be $o\in
M_\beta^{(2)}$ such that $(n,o,q)\notin \CA_{\alpha,\beta,\gamma,j}$ while
\eqref{E:o_cond} and \eqref{E:n_cond} are satisfied. This is a contradiction to
the definition of $\CA_{\alpha,\beta,\gamma,j}$. After that we sum over all
$q\in M_\gamma$ and the result follows from \eqref{E:NL_term_conv}.
\erem%

We now write the $d$-th component ($d\in \{1,2,3\}$) of $\wt{G}_j$ as
\begin{align}
  \label{E:Gamma_def}
  \wt{G}_{j,d}(k;x)=\eps^{-4}\chi_{D_{\eps^r}}(k-k^{(j)}+K^m)
  \sum_{a,b,c=1}^3\Gamma_{a,b,c}^{(d)}
  \left(\wt{E}^{(0)}_a *_{\Bsm} \wt{E}^{(0)}_b *_{\Bsm}
  \wt{\barl{E^{(0)}_c}}\right)(k;x),
\end{align}
where the integer coefficients $\Gamma_{a,b,c}^{(d)}$ can be easily derived
from \eqref{E:PNL_ans_full}. In detail we have
$\Gamma_{1,1,1}^{(1)}=\Gamma_{2,2,2}^{(2)}=\Gamma_{3,3,3}^{(3)}=3$,
$\Gamma_{1,2,2}^{(1)}=\Gamma_{2,1,2}^{(1)}=\Gamma_{1,3,3}^{(1)}
 =\Gamma_{3,1,3}^{(1)}
 =\Gamma_{2,2,1}^{(1)}=\Gamma_{3,3,1}^{(1)}=1$,
$\Gamma_{1,2,1}^{(2)}=\Gamma_{2,1,1}^{(2)}=\Gamma_{3,2,3}^{(2)}
 =\Gamma_{2,3,3}^{(2)}=\Gamma_{1,1,2}^{(2)}=\Gamma_{3,3,2}^{(2)}=1$,
$\Gamma_{1,3,1}^{(3)}=\Gamma_{3,1,1}^{(3)}=\Gamma_{2,3,2}^{(3)}
 =\Gamma_{3,2,2}^{(3)}=\Gamma_{1,1,3}^{(3)}=\Gamma_{2,2,3}^{(3)}=1$,
and the remaining $\Gamma_{a,b,c}^{(d)}$ are zero. Finally, using
\eqref{E:NL_term_conv_sum}, we get for $k\in B_{\eps^r}(k^{(j)}-K^m)$
\begin{align}\label{E:NL_term_structure}
\begin{split}
  \wt{G}_{j,d}(k;x)=e^{\ri(-k^{(j)}+K^m)\cdot x}
   \sum_{a,b,c=1}^3\Gamma_{a,b,c}^{(d)}
   \sum_{\substack{\alpha,\beta,\gamma\in \{1,\ldots,N\} \text{ s.t.} \\
   \CA_{\alpha,\beta,\gamma,j}\neq \emptyset}}
  & u_{n_\alpha,a}(k^{(\alpha)};x)u_{n_\beta,b}(k^{(\beta)};x)\\
   \times & \barl{u_{n_\gamma,c}}(k^{(\gamma)};x)
   \tilde{h}_{\alpha,\beta,\gamma}^{(\eps)}(\ell^{(j,m)}(k)).
\end{split}
\end{align}

In order to make the discussion of the asymptotic hierarchy complete, we also
have to consider the part of the $k-$domain outside the neighborhoods of
$k^{(j)}$. For $k \in \B$ such that $k-k^{(j)}+K^m \in \B\setminus D_{\eps^r}$
for all $m \in M_j$ we have
$\bigl(\wt{L}(k^{(j)}-K^m;x)-\omega_*^2\eta(x)\bigr)\wt{E}^{(l)}(k;x)=0$ for
$l\in\{0,1\}$ so that $\wt{E}^{(0)}(k;\pkt)\equiv \wt{E}^{(1)}(k;\pkt)\equiv 0$
for such $k$.

\subsection{Coupled mode equations}%
\label{S:cme}%

\begin{figure}[h!]
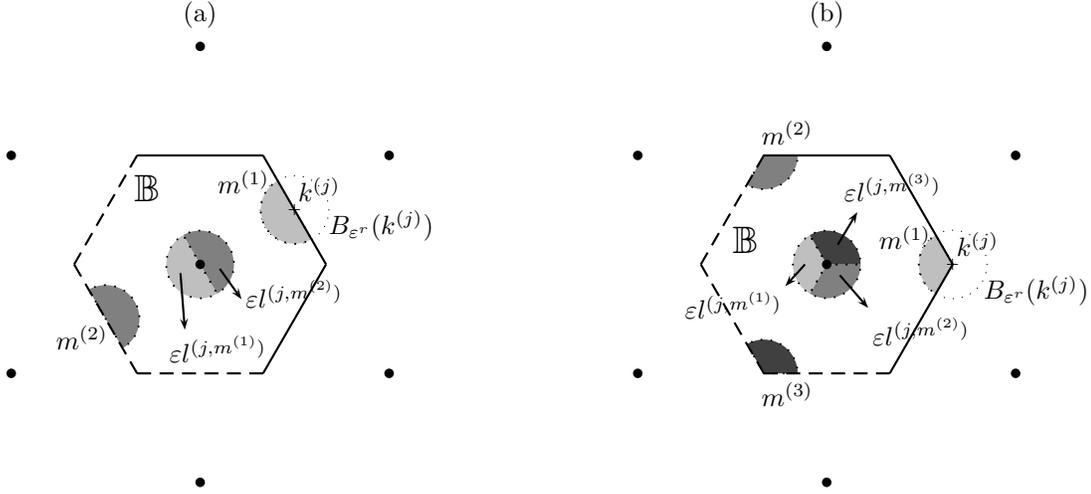

\begin{center}
 \begin{minipage}[c]{.49\textwidth}
 \centering
 (a)

 \medskip
 \psset{unit=2.9cm}
 \pspicture(-1,-1)(1,1)

 \psdots*(0,0)(0.866,0.5)(0.866,-0.5)(-0.866,0.5)(-0.866,-0.5)(0,1)(0,-1)
 \put(-0.3,0.29){\Large$\B$}
 \pscircle[linestyle=dotted,linewidth=0.5pt](0.433,0.25){0.16}
 \pswedge[linestyle=dotted,fillstyle=solid,fillcolor=gray](-0.433,-0.25){0.16}{-60}{120}
 \pswedge[linestyle=dotted,fillstyle=solid,fillcolor=lightgray](0.433,0.25){0.16}{120}{300}
 \psline(0.5774,0)(0.2887,0.5)
 \psline(0.2887,0.5)(-0.2887,0.5)
 \psline[linestyle=dashed](-0.2887,0.5)(-0.5774,0)
 \psline[linestyle=dashed](-0.5774,0)(-0.2887,-0.5)
 \psline[linestyle=dashed](-0.2887,-0.5)(0.2887,-0.5)
 \psline(0.2887,-0.5)(0.5774,0)
 \psdots*[dotstyle=+](0.433,0.25)
 \put(0.45,0.29){$k^{(j)}$}
 \put(0.59,0.14){$B_{\eps^r}(k^{(j)})$}

 \pswedge[linestyle=dotted,fillstyle=solid,fillcolor=gray](0,0){0.16}{-60}{120}
 \pswedge[linestyle=dotted,fillstyle=solid,fillcolor=lightgray](0,0){0.16}{120}{300}
 \psdots*(0,0)
 \put(0.08,0.325){$m^{(1)}$}
 \put(-0.66,-0.39){$m^{(2)}$}
 \psline{->}(0.09,-0.02)(0.19,-0.16)
 \psline{->}(-0.09,-0.04)(-0.07,-0.3)
 \put(0.21,-0.2){$\eps l^{(j,m^{(2)})}$}
 \put(-0.14,-0.45){$\eps l^{(j,m^{(1)})}$}
 \endpspicture
 \end{minipage}
 \begin{minipage}[c]{0.49\textwidth}
 \centering
 (b)

 \medskip
 \psset{unit=2.9cm}
 \pspicture(-1,-1)(1,1)
 \psdots*(0,0)(0.866,0.5)(0.866,-0.5)(-0.866,0.5)(-0.866,-0.5)(0,1)(0,-1)
 \put(-0.43,0.05){\Large$\B$}
 \pscircle[linestyle=dotted,linewidth=0.5pt](0.5774,0){0.16}
 \pswedge[linestyle=dotted,fillstyle=solid,fillcolor=lightgray](0.5774,0){0.16}{120}{240}
 \pswedge[linestyle=dotted,fillstyle=solid,fillcolor=gray](-0.2887,0.5){0.16}{240}{360}
 \pswedge[linestyle=dotted,fillstyle=solid,fillcolor=darkgray](-0.2887,-0.5){0.16}{0}{120}
 \psline(0.5774,0)(0.2887,0.5)
 \psline(0.2887,0.5)(-0.2887,0.5)
 \psline[linestyle=dashed](-0.2887,0.5)(-0.5774,0)
 \psline[linestyle=dashed](-0.5774,0)(-0.2887,-0.5)
 \psline[linestyle=dashed](-0.2887,-0.5)(0.2887,-0.5)
 \psline(0.2887,-0.5)(0.5774,0)
 \psdots*[dotstyle=+](0.5774,0)
 \put(0.595,0.04){$k^{(j)}$}
 \put(0.72,-0.16){$B_{\eps^r}(k^{(j)})$}

 \put(0.24,0.07){$m^{(1)}$}
  \put(-0.3,0.55){$m^{(2)}$}
 \put(-0.3,-0.65){$m^{(3)}$}
 \pswedge[linestyle=dotted,fillstyle=solid,fillcolor=lightgray](0,0){0.16}{120}{240}
 \pswedge[linestyle=dotted,fillstyle=solid,fillcolor=gray](0,0){0.16}{240}{360}
 \pswedge[linestyle=dotted,fillstyle=solid,fillcolor=darkgray](0,0){0.16}{0}{120}
 \psdots*(0,0)
 \psline{->}(0.06,-0.05)(0.19,-0.2)
 \psline{->}(0.05,0.09)(0.14,0.24)
 \psline{->}(-0.1,0)(-0.19,-0.1)
 \put(0.21,-0.37){$\eps l^{(j,m^{(2)})}$}
 \put(-0.65,-0.25){$\eps l^{(j,m^{(1)})}$}
 \put(0.08,0.29){$\eps l^{(j,m^{(3)})}$}
 \endpspicture
 \end{minipage}
\end{center}
\caption{\label{F:Dj_Mj}\small%
Two example points $k^{(j)}$ in the case of the hexagonal lattice and the
corresponding ranges of $\eps\ell^{(j,m)}$ for all $m\in M_j$. In (a) we have
$M_j=\{(0,0)^T,(1,1)^T\}=:\{m^{(1)},m^{(2)}\}$ and in (b) $M_j=\{(0,0)^T,
(0,1)^T, (1,1)^T\}=:\{m^{(1)},m^{(2)},m^{(3)}\}$. The shaded sections along the
boundary of $\B$ are those $k\in \B$ for which
$\chi_{D_{\eps^r}}(k-k^{(j)}+K^m)\neq 0$ for the $m\in M_j$ written next to the
respective section.}
\end{figure}

We return now to equation \eqref{E:Oeps}. Due to the Fredholm alternative the
existence of $\Lambda$-periodic solutions $\wt{E}^{(2)}$ of equation
\eqref{E:Oeps} is equivalent to $L^2$-orthogonality of \eqref{E:Oeps} to
$p_{n_j}(k^{(j)};x)e^{\ri K^m\cdot x}$, which needs to be ensured for all $m
\in M_j$ and $j\in \{1,\ldots, N\}$. The range of $\ell^{(j,m)}$ is a different
section of the disc $D_{\eps^{r-1}}$ for each $m\in M_j$. This section is a
$(1/|M_j|)$-th of the full disc so that these $|M_j|$ equations actually build
one equation in $\ell\in D_{\eps^{r-1}}$. Figure \ref{F:Dj_Mj} shows these
sections for two example points $k^{(j)}$. One example is for $|M_j|=2$ and the
other one for $|M_j|=3$.

When imposing the orthogonality condition, the common factor $e^{\ri K^m\cdot
x}$ of the right hand side of \eqref{E:Oeps} is canceled in the complex inner
product with $p_{n_j}(k^{(j)};x)e^{\ri K^m\cdot x}$, so that the same
solvability condition holds for all $m \in M_j$. Using \eqref{E:Oeps},
\eqref{E:normalize_Bloch}, and \eqref{E:om_deriv_2} (with $n_*$ and $k_*$
replaced by $n_j$ and $k^{(j)}$), we obtain
\beq\label{E:CME_Four}%
  \Omega\wh{A}_j(\ell)
   -\frac{1}{2}\left(\ell_1^2\pa_{k_1}^2\omega_{n_j}(k^{(j)})
   +\ell_2^2\pa_{k_2}^2\omega_{n_j}(k^{(j)})
   +2\ell_1\ell_2\pa_{k_1,k_2}^2\omega_{n_j}(k^{(j)})\right)\wh{A}_j(\ell)
   + \wh{\CN}_j(\ell) = 0
\eeq%
for $\ell\in D_{\eps^{r-1}}$, where
\begin{align*}
   \wh{\CN}_j(\ell)
   &=\frac{\omega_*}{2}\big\langle\chici(\pkt)\wt{G}_j(\ell;\pkt),
     p_{n_j}(k^{(j)};\pkt)e^{-\ri K^m\cdot(\pkt)}\big\rangle\\
   &=\frac{\omega_*}{2}\sum_{a,b,c,d=1}^3\Gamma_{a,b,c}^{(d)}
    \sum_{\substack{\alpha,\beta,\gamma\in \{1,\ldots,N\} \text{ s.t.}\\
   \CA_{\alpha,\beta,\gamma,j}\neq \emptyset}}\int_U\chici(x)
     u_{n_\alpha,a}(k^{(\alpha)};x)u_{n_\beta,b}(k^{(\beta)};x)\\
    & \hspace{6.5cm}\times \barl{u_{n_\gamma,c}}(k^{(\gamma)};x)
     \barl{u_{n_j,d}}(k^{(j)};x)\dd x\;
     \tilde{h}_{\alpha,\beta,\gamma}^{(\eps)}(\ell)\\
   &=:\sum_{\substack{\alpha,\beta,\gamma\in \{1,\ldots,N\} \text{ s.t.}\\
    \CA_{\alpha,\beta,\gamma,j}\neq \emptyset}}
     I_{\alpha,\beta,\gamma,j}\tilde{h}_{\alpha,\beta,\gamma}^{(\eps)}(\ell),
\end{align*}
i.e. with \eqref{E:PNL_ans_full} and the definition of $\Gamma$ in \eqref{E:Gamma_def}
\beq \label{E:I_coeffs}
 \begin{array}{rl}
   I_{\alpha,\beta,\gamma,j}
   :=& \displaystyle
      \frac{\omega_*}{2}\sum_{a,b,c,d=1}^3\Gamma_{a,b,c}^{(d)} \int_U\chici
      u_{n_\alpha,a}(k^{(\alpha)};\pkt)u_{n_\beta,b}(k^{(\beta)};\pkt)
    \barl{u_{n_\gamma,c}}(k^{(\gamma)};\pkt)\barl{u_{n_j,d}}(k^{(j)};\pkt)\\
    =& \displaystyle
      \frac{\omega_*}{2} \int_U \chici
      \left[2(u_{n_\alpha}(k^{(\alpha)};\pkt)
      \cdot \barl{u_{n_\gamma}}(k^{(\gamma)};\pkt))
      u_{n_\beta}(k^{(\beta)};\pkt) \right.\\
     &\left. \qquad\; + (u_{n_\alpha}(k^{(\alpha)};\pkt)
      \cdot u_{n_\beta}(k^{(\beta)};\pkt))
      \barl{u_{n_\gamma}}(k^{(\gamma)};\pkt)\right]\cdot\barl{u_{n_j}}(k^{(j)};\pkt)\;.
 \end{array}
\eeq%
The symmetries in $\Gamma_{a,b,c}^{(d)}$ imply symmetries in
$I_{\alpha,\beta,\gamma,j}$. Namely, due to the symmetries
$\Gamma_{a,b,c}^{(d)}=\Gamma_{b,a,c}^{(d)}$ and
$\Gamma_{a,b,c}^{(d)}=\Gamma_{a,b,d}^{(c)}$ we have
\beq \label{E:I_sym1}%
   I_{\alpha,\beta,\gamma,j}=I_{\beta,\alpha,\gamma,j} \text{ and }
   I_{\alpha,\beta,\gamma,j}=I_{\alpha,\beta,j,\gamma} \qquad\text{for all }
    \alpha,\beta,\gamma,j\in\{1,\ldots,N\},
\eeq%
and due to $\Gamma_{a,b,c}^{(d)}=\Gamma_{c,d,a}^{(b)}$ we have
\beq\label{E:I_sym2}%
   I_{\alpha,\beta,\gamma,j}=\overline{I_{\gamma,j,\alpha,\beta}}
    \qquad\text{for all } \alpha,\beta,\gamma,j\in\{1,\ldots,N\}.
\eeq%
Symmetries \eqref{E:I_sym1} and \eqref{E:I_sym2} imply, in particular, that
$I_{\alpha,\beta,\alpha,\beta}=I_{\alpha,\beta,\beta,\alpha}\in\mathbb\R$ for
all $\alpha,\beta\in\{1,\ldots,N\}$.

Let the crystal satisfy the rotational symmetry $\eta(x)=\eta(r_\nu(x))$ and
$\chici(x)=\chici(r_\nu(x))$ for all $x\in \R^2$ and some $\nu\in (-\pi,\pi]$
and let $U$ be chosen so that $r_\nu(U)=U$. If for each $m\in\{\alpha,\beta,\gamma,j\}\subset\{1,\dots,N\}$
there exists $m'\in\{1,\dots,N\}$ such that
\[
   k^{(m')}=r_\nu(k^{(m)}),
\]%
and if $\omega_n(k^{(m)})$ is a geometrically simple eigenvalue of
\eqref{E:shifted} for each $m\in \{\alpha,\beta,\gamma,j\}$, then
\beq\label{E:I_sym_rot}%
   I_{\alpha,\beta,\gamma,j}=I_{\alpha',\beta',\gamma',j'}.
\eeq%
This is seen by the change of variables $y=r_\nu(x)$ in \eqref{E:I_coeffs},
using the facts $r_\nu(U)=U$ and $r_\nu(v)\cdot r_\nu(w)=v\cdot w$ for all
$v,w\in \C^3$, and employing the symmetry \eqref{E:rot_sym_Bloch}.

Additional symmetries in $I_{\alpha,\beta,\gamma,j}$ arise when a spatial
reflection symmetry in $\eta$ and $\chici$ is present. For instance when
$\eta(x)=\eta(S_2(x))$, $\chici(x)=\chici(S_2(x))$ for all $x\in \R^2$ (see
Section \ref{S:refl_sym}) and if for each $m\in\{\alpha,\beta,\gamma,j\}\subset\{1,\dots,N\}$ there exists
$m'\in\{1,\dots,N\}$ such that
\[
   k^{(m')} = S_2(k^{(m)})
\]%
and such that $S_2(k^{(m)}) \doteq k^{(m)}$ does \textit{not} hold for
any $m\in \{\alpha,\beta,\gamma,j\}$, then
\beq\label{E:I_sym_refl}%
   I_{\alpha,\beta,\gamma,j} = I_{\alpha',\beta',\gamma',j'}.
\eeq%
This is proved via a change of variables in \eqref{E:I_coeffs} using
\eqref{E:refl2_sym_Bloch}, where $a=0$ due to our assumptions. A similar result
holds for the reflection symmetry in $x_1$.

Returning now back to \eqref{E:CME_Four}, for smooth envelopes $A_j$ we can
neglect the contribution of $\wh{A}_j$ from $\ell\in \R^2\setminus
D_{\eps^{r-1}}$ or simply assume that $\wh{A}_j$ satisfy \eqref{E:CME_Four}
also there. This step can be rigorously justified via a persistence argument
similar to that in \cite{DU09,DU_err11}.
$\tilde{h}_{\alpha,\beta,\gamma}^{(\eps)}$ will then be replaced by
$\wh{A}_{\alpha}*\wh{A}_{\beta}*\wh{\barl{A}}_{\gamma}$. The inverse Fourier
transform then produces the \emph{coupled mode equations}
\beq\label{E:CME}%
   \Omega A_j
   +\frac{1}{2}\left(\pa_{k_1}^2\omega_{n_j}(k^{(j)})\pa_{y_1}^2
   +\pa_{k_2}^2\omega_{n_j}(k^{(j)})\pa_{y_2}^2
   +2\pa^2_{k_1,k_2}\omega_{n_j}(k^{(j)})\pa_{y_1,y_2}^2\right)A_j
   +\CN_j =0
\eeq%
on $\R^2$, where $\CN_j$ is given by
\begin{align*}
   \CN_j
   &=\sum_{\substack{\alpha,\beta,\gamma\in \{1,\ldots,N\} \text{ s.t.}\\
     \CA_{\alpha,\beta,\gamma,j}\neq \emptyset}}
     I_{\alpha,\beta,\gamma,j}A_\alpha A_\beta\barl{A_\gamma}.
\end{align*}
Note that the coupled mode equations have the same general structure as those
for gap solitons of the scalar Gross--Pitaevskii equation \cite{DU09}.

A localized solution $A$ of \eqref{E:CME} should produce via
\eqref{E:ansatz_phys} an approximation of a gap soliton of the Maxwell problem
\eqref{E:NL_Maxw}. A rigorous justification of this statement can be done via
the Lyapunov--Schmidt reduction similarly to \cite{DPS09,DU09,DU_err11} and
will be the subject of a future project. System \eqref{E:CME} does not have
localized solutions for arbitrary values of coefficients. The coefficients of
the derivative terms are given by the band structure and $\Omega=\pm 1$ is
determined by the condition that $\omega=\omega_* + \eps^2 \Omega$ lies in the
gap. But the function $\chici$ in $I_{\alpha,\beta,\gamma,j}$ has not been
fixed and remains free at this point.

The linear part of the operator in \eqref{E:CME} is definite due to our
assumption (A4) in Section \ref{S:slowly envelope} and the fact that $\Omega<0$
at upper edges and $\Omega>0$ at lower edges. The linear part of the operator
is positive definite at lower edges $\omega_*$, where $k^{(j)}$ are points of
maxima and negative definite at upper edges. In case $N=1$, where $\CN_1
=\gamma |A_1|^2A_1$ and $\gamma =\tfrac{3\omega_*}{2}\int_U
\chici|u_{n_j}(k^{(1)};\pkt)|^4$, a localized solution exists in the upper edge
case only if $\chici$ is such that $\gamma>0$ while in the lower edge case
$\chici$ has to produce $\gamma<0$. Physically it makes sense to set $\chici=0$
there, where $\eta=1$ (i.e.{} in vacuum/air). In the annulus regions, where
$\eta=2.1025$, we set $\chici=1$ (a focusing nonlinearity)  if $\gamma>0$ is
needed and $\chici=-1$ (a defocusing nonlinearity) if $\gamma<0$ is required.
This is in agreement with previous results on bifurcation of gap solitons from
spectral edges in the periodic nonlinear Schr\"odinger equation
\cite{HKS92,AL92,Pankov05,DPS09,DU09}, where bifurcation from upper/lower edges
occurs for the focusing/defocusing nonlinearity respectively. In the case $N>1$
our numerical examples produce all $I_{\alpha,\beta,\gamma,j}$ of the same sign
so that we set in the annulus regions, once again, $\chici=1$ if $\omega_*$ is
an upper edge of a gap and $\chici=-1$ if it is a lower edge.

\subsection{Examples of Coupled Mode Equations}\label{S:CME_examples}

We present next coupled mode equations for gap solitons in the vicinity of
spectral edges for the example in Section \ref{S:hex_struct} as well as for
other canonical examples. As seen in Figure \ref{F:band_str}, there are 3
spectral gaps $(0,s_1)$, $(s_2,s_3)$ and $(s_4,s_5)$ on the positive part of
the spectral $\omega$ axis for this specific example. We have the numerical
values
\[
 \begin{array}{c}
   s_1=\omega_1(\Gamma)\approx 3.610,\;
   s_2=\omega_6(\Gamma)\approx 3.701,\;
   s_3=\omega_7(\Gamma)\approx 3.750,\\
   s_4=\omega_{12}(0,2.351)\approx 3.873,\;
   s_5=\omega_{13}(0,2.407)\approx 3.882.
 \end{array}
\]
At $s_1$ and $s_2$ several bands lie very close to each other at the extremal
point $k=\Gamma$. It is, however, not known whether these truly touch and
the eigenvalues have higher multiplicity than one. Numerical tests have
shown that varying the value of $\eta$ for the annulus material does not change
the ordering of bands at $k=\Gamma$ near $s_1$ and $s_2$. We, therefore, assume
that the edges $s_1$ and $s_2$ are simple eigenvalues at $k=\Gamma$ leading to
$N=1$ at $s_1$ and $s_2$. If it can be proved that, for instance, $s_1$ is
indeed a double eigenvalue, then $N=2$ at $s_1$. Likewise, $N$ would change if
the multiplicity could be established for $s_2$.

Similarly, the band $\omega_{12}$ is close to $\omega=s_5$ at four distinct
$k$-points along $\pa \B_0$. At the point $k=(0,2.351)$ the numerical value is
maximal and an analogous test shows that it remains maximal for a range of
values of $\eta$. We thus assume that within $\B_0$ the value $\omega=s_5$ is
attained only at $k=(0,2.351)$. Due to the discrete rotational symmetry of the
band structure we thus have $N=6$ at $s_5$. Analogously, we have $N=6$ at
$s_4$.

Except for the simplest case with $N=1$, like in Section \ref{S:CME_N1}, we
determine the sets $\CA_{\alpha,\beta,\gamma,j}$ using a Matlab program. First
of all, it is clear that for any $k^{\flat}\in\B$ the sets $M_\flat$ and
$M_\flat^{(2)}$ contain only those $(n,o,q)\in \Z^2$ with $n_l,o_l,q_l\in
\{-1,0,1\}$ for $l=1,2$. To determine $\CA_{\alpha,\beta,\gamma,j}$, we
therefore need to test only finitely many integer vectors $(n,o,q)$ for
conditions \eqref{E:o_cond}, \eqref{E:n_cond}. For an example with $N=3$ we
show in Section \ref{S:CME_3_extrema} the resulting sets
$\CA_{\alpha,\beta,\gamma,j}$ computed using this routine.

\subsubsection{Coupled Mode Equations near Edges for the Example in
Section \ref{S:hex_struct}}\label{S:CME_hex}%

\paragraph{Coupled Mode Equations near the Edges $s_1,s_2$ and $s_3$ ($N=1$)}
\label{S:CME_N1}\ %

At the edges $s_1$, $s_2$ and $s_3$ in Figure \ref{F:band_str} the situation is
particularly simple. As discussed at the beginning of Section
\ref{S:CME_examples}, we have $N=1$ and $k^{(1)} =\Gamma =\bspm 0 \\0 \espm$.
Since $k^{(1)}\in \operatorname{int}(\B)$, any (small) neighborhood of
$k^{(1)}$ lies completely within $\B$ and thus $M_1=\{\bspm 0 \\0 \espm\}$. A
simple inspection determines that we have $\CA_{1,1,1,1} = \big\{(\bspm 0 \\0
\espm, \bspm 0\\0 \espm, \bspm 0 \\0 \espm)\big\}$. The resulting coupled mode
equation for $A=A_1$ is
\beq\label{E:CME_1}%
   \left(\Omega +\alpha (\partial_{y_1}^2
   +\partial_{y_2}^2)\right)A+\gamma |A|^2A=0,
\eeq%
where $\alpha =\tfrac{1}{2}\partial_{k_1}^2\omega_{n_1}(\Gamma)
=\tfrac{1}{2}\partial_{k_2}^2\omega_{n_1}(\Gamma)$ (cf.{} \eqref{E:der2_Gamma})
and $\gamma=I_{1,1,1,1}$.

The three cases $s_1,s_2$ and $s_3$ differ by the value of $n_1$, i.e.{} the
band index. At $\omega_*=s_1$ we have $n_1=1$, at $\omega_*=s_2$ we have
$n_1=6$ and at $\omega_*=s_3$ we have $n_1=7$. And, as discussed at the end of
Section \ref{S:cme}, at the upper edges $s_1,s_3$ we have $\Omega=-1$ and the
function $\chici$ has the value $1$ in the annulus regions and $0$ otherwise.
At $s_2$ we have $\Omega=1$ and $\chici=-1$ in the annuli.

In Section \ref{S:numerics_s2} we present  a numerical example on a gap soliton
approximation near $s_2$. We list here, therefore, the numerical values of the
CME coefficients for the case $s_2$:
\[
   \omega_*=s_2\approx 3.701: \alpha\approx -0.0107,\ \gamma\approx -3.057.
\]

\paragraph{Coupled Mode Equations near the Edge $s_5$ ($N=6$)}\ %

At the upper edge $s_5$ in Figure \ref{F:band_str} we have $N=6$,
$n_1=n_2=\ldots=n_6=13$, $k^{(1)} \approx (0,2.458)$ lying on the line from
$\Gamma$ to $r_{2\pi/3}(M)$, and $k^{(j)}$, $j=2,\ldots,6$, obtained via a
rotation of $k^{(1)}$. In detail
\[
  k^{(j)} = r_{(j-1)\tfrac{\pi}{3}}(k^{(1)}) \qquad\text{for }j=2,\ldots,6.
\]
The symmetry properties \eqref{E:der2_rot_sym}, \eqref{E:der2_refl_sym}, and
\eqref{E:der2_refl_symB} produce relations among the linear coefficients of the
CMEs. The sets $\CA_{\alpha,\beta,\gamma,j}$ are either empty or contain solely
the element $\{\bspm 0 \\0 \espm, \bspm 0 \\0 \espm, \bspm 0 \\0 \espm\}$ as
checked by the Matlab routine. The resulting CMEs are
\beq\label{E:CME_s5}
\begin{array}{rl}
 (\Omega+\alpha_1\pa_{y_1}^2+\beta_1\pa_{y_2}^2)A_1+N_1 =&0,\\
 (\Omega+\alpha_2\pa_{y_1}^2+\beta_2\pa_{y_2}^2 +\mu\pa^2_{y_1,y_2})A_2+N_2 =& 0,\\
 (\Omega+\alpha_2\pa_{y_1}^2+\beta_2\pa_{y_2}^2 -\mu\pa^2_{y_1,y_2})A_3+N_3 =& 0,\\
 (\Omega+\alpha_1\pa_{y_1}^2+\beta_1\pa_{y_2}^2)A_4+N_4 =& 0,\\
 (\Omega+\alpha_2\pa_{y_1}^2+\beta_2\pa_{y_2}^2 +\mu\pa^2_{y_1,y_2})A_5+N_5 =& 0,\\
 (\Omega+\alpha_2\pa_{y_1}^2+\beta_2\pa_{y_2}^2 -\mu\pa^2_{y_1,y_2})A_6+N_6 =& 0,
\end{array}
\eeq%
where $\Omega=-1$, $\alpha_1 = \pa_{k_1}^2\omega_{13}(k^{(1)})$, $\beta_1 =
\pa_{k_2}^2\omega_{13}(k^{(1)})$, $\alpha_2 = \tfrac{1}{4}(\alpha_1+3\beta_1)$,
$\beta_2=\tfrac{1}{4}(3\alpha_1+\beta_1)$, $\mu =
\tfrac{\sqrt{3}}{4}(\alpha_1-\beta_1) = \pa_{k_1,k_2}^2\omega_{13}(k^{(2)})$,
and
\[
 \begin{split}
  N_1 & = 2\sum_{i=1}^6 I_{i,1,i,1}|A_i|^2A_1-I_{1,1,1,1}|A_1|^2A_1
  +2(I_{2,5,4,1}A_2A_5+I_{3,6,4,1}A_3A_6)\bar{A}_4,\\
  N_2 & = 2\sum_{i=1}^6 I_{i,2,i,2}|A_i|^2A_2-I_{2,2,2,2}|A_2|^2A_2
  +2(I_{1,4,5,2}A_1A_4+I_{3,6,5,2}A_3A_6)\bar{A}_5,\\
  N_3 & = 2\sum_{i=1}^6 I_{i,3,i,3}|A_i|^2A_3-I_{3,3,3,3}|A_3|^2A_3
  +2(I_{1,4,6,3}A_1A_4+I_{2,5,6,3}A_2A_5)\bar{A}_6,\\
  N_4 & = 2\sum_{i=1}^6 I_{i,4,i,4}|A_i|^2A_4-I_{4,4,4,4}|A_4|^2A_4
  +2(I_{2,5,1,4}A_2A_5+I_{3,6,1,4}A_3A_6)\bar{A}_1,\\
  N_5 & = 2\sum_{i=1}^6 I_{i,5,i,5}|A_i|^2A_5-I_{5,5,5,5}|A_5|^2A_5
  +2(I_{1,4,2,5}A_1A_4+I_{3,6,2,5}A_3A_6)\bar{A}_2,\\
  N_6 & = 2\sum_{i=1}^6 I_{i,6,i,6}|A_i|^2A_6-I_{6,6,6,6}|A_6|^2A_6
  +2(I_{1,4,3,6}A_1A_4+I_{2,5,3,6}A_2A_5)\bar{A}_3.
\end{split}
\]%
Due to symmetries, many of the coefficients in the nonlinear terms are equal.
Symmetry \eqref{E:I_sym_rot} with $\nu=\pi/3$ and symmetry \eqref{E:I_sym1}
imply
\[
 \begin{split}
  \gamma_0:=&I_{1,1,1,1}= I_{2,2,2,2}=\ldots=I_{6,6,6,6},\\
  \gamma_1:=&I_{2,1,2,1}=I_{3,2,3,2}=\ldots=I_{6,5,6,5}=I_{1,6,1,6}\\
  =&I_{1,2,1,2}=I_{2,3,2,3}=\ldots=I_{5,6,5,6}=I_{6,1,6,1},\\
  \gamma_2:=&I_{3,1,3,1}=I_{4,2,4,2}=I_{5,3,5,3}=I_{6,4,6,4}=I_{1,5,1,5}=I_{2,6,2,6}\\
  =&I_{1,3,1,3}=I_{2,4,2,4}=I_{3,5,3,5}=I_{4,6,4,6}=I_{5,1,5,1}=I_{6,2,6,2},\\
  \gamma_3:=&I_{4,1,4,1}=I_{5,2,5,2}=I_{6,3,6,3}
     =I_{1,4,1,4}=I_{2,5,2,5}=I_{3,6,3,6},\\
  \gamma_4:=&I_{2,5,1,4}=I_{3,6,2,5}=I_{1,4,3,6}
     =I_{2,5,4,1}=I_{3,6,5,2}=I_{1,4,6,3}.
 \end{split}
\]
Using \eqref{E:I_sym2} and \eqref{E:I_sym1}, we get
\[
   \overline{\gamma_4}=I_{3,6,4,1}=I_{3,6,1,4}=I_{1,4,5,2}=I_{1,4,2,5}
   =I_{2,5,6,3}=I_{2,5,3,6}.
\]%
We have $\gamma_0,\gamma_1,\gamma_2,\gamma_3\in \R$ as explained below
\eqref{E:I_sym2}.

Finally, because $k^{(1)}=(k^{(4)}_1,-k^{(4)}_2)^T$,
$k^{(2)}=(k^{(3)}_1,-k^{(3)}_2)^T$, and $k^{(5)}=(k^{(6)}_1,-k^{(6)}_2)^T$ with
$k^{(1)}, k^{(2)}$ and $k^{(5)}$ lying in the interior of $\B$ away from the
line $k_2=0$, the symmetry \eqref{E:I_sym_refl} applies and we get
\[
   I_{2,5,4,1}=I_{3,6,1,4}.
\]%
Therefore
\beq\label{E:I_real}
   I_{2,5,4,1}=I_{3,6,1,4}=I_{3,6,4,1}=I_{4,1,5,2}=\overline{I_{2,5,4,1}}
\eeq%
so that also $\gamma_4\in\R$. The second, third and fourth equalities in
\eqref{E:I_real} hold due to \eqref{E:I_sym1}, \eqref{E:I_sym_rot}, and
\eqref{E:I_sym2}.

As a result the nonlinear terms in \eqref{E:CME_s5} can be simplified to
\[
 \begin{split}
  N_1 & := 2\left(\frac{\gamma_0}{2}|A_1|^2+\gamma_1(|A_2|^2+|A_6|^2)+\gamma_2(|A_3|^2+|A_5|^2)
     +\gamma_3|A_4|^2\right)A_1 +2\gamma_4(A_2A_5+A_3A_6)\barl{A_4},\\
  N_2 & := 2\left(\frac{\gamma_0}{2}|A_2|^2+\gamma_1(|A_1|^2+|A_3|^2)+\gamma_2(|A_4|^2+|A_6|^2)
     +\gamma_3|A_5|^2\right)A_2 +2\gamma_4(A_1A_4+A_3A_6)\barl{A_5},\\
  N_3 & := 2\left(\frac{\gamma_0}{2}|A_3|^2+\gamma_1(|A_2|^2+|A_4|^2)+\gamma_2(|A_1|^2+|A_5|^2)
     +\gamma_3|A_6|^2\right)A_3 +2\gamma_4(A_1A_4+A_2A_5)\barl{A_6},\\
  N_4 & := 2\left(\frac{\gamma_0}{2}|A_4|^2+\gamma_1(|A_3|^2+|A_5|^2)+\gamma_2(|A_2|^2+|A_6|^2)
     +\gamma_3|A_1|^2\right)A_4 +2\gamma_4(A_2A_5+A_3A_6)\barl{A_1},\\
  N_5 & := 2\left(\frac{\gamma_0}{2}|A_5|^2+\gamma_1(|A_4|^2+|A_6|^2)+\gamma_2(|A_1|^2+|A_3|^2)
     +\gamma_3|A_2|^2\right)A_5 +2\gamma_4(A_1A_4+A_3A_6)\barl{A_2},\\
  N_6 & := 2\left(\frac{\gamma_0}{2}|A_6|^2+\gamma_1(|A_1|^2+|A_5|^2)+\gamma_2(|A_2|^2+|A_4|^2)
     +\gamma_3|A_3|^2\right)A_6 +2\gamma_4(A_1A_4+A_2A_5)\barl{A_3}
 \end{split}
\]
with $\gamma_0,\gamma_1,\gamma_2,\gamma_3,\gamma_4\in \R$. A system of six CMEs
with the same structure as above arises also at the edge $s_4$.

In Section \ref{S:numerics_s5} a numerical example of gap soliton asymptotics
near $s_5$ is given. The numerical values of the coefficients in the CMEs
\eqref{E:CME_s5} for $s_5$ are
\[
 \begin{split}
    \omega_*=s_5 \approx 3.882:\
    & \alpha_1 \approx 0.0189,\,\alpha_2 \approx 0.146,\,\beta_1 \approx 0.189,\,
      \beta_2 \approx 0.0614,\,\mu\approx -0.0736,\\
    & \gamma_0\approx 1.282,\,\gamma_1\approx 0.789, \gamma_2\approx 0.757,\,
      \gamma_3\approx 1.193,\,\gamma_4\approx 0.714.
 \end{split}
\]%
As $s_5$ is an upper edge edge, the coefficients $\gamma_j$,
$j\in\{0,\dots,4\}$, were computed using $\chici=1$ in the annulus regions.


\subsubsection{Additional CME Examples}

\paragraph{Example of Coupled Mode Equations for $N=2$}\label{S:CME_N2}\ %

An example of a  situation for $N=2$ is when the locations of the extrema are
$k^{(1)}=K$, $k^{(2)}=r_{\pi/3}(K)$. With $b^{(1)}, b^{(2)}$ as in Section
\ref{S:hex_struct} we then have $k^{(1)} =\tfrac{4\pi}{3a_0}\bspm 1 \\0 \espm$
and $k^{(2)} =\tfrac{2\pi}{3a_0}\bspm 1\\\sqrt{3} \espm$.
The corresponding integer shift sets are $M_1 = \{\bspm 0 \\0 \espm, \bspm 0\\
1 \espm, \bspm 1 \\1 \espm\}$, $M_2 = \{\bspm 0 \\0 \espm, \bspm 1 \\0 \espm,
\bspm 1 \\1 \espm\}$. Due to the rotation symmetry of the bands and their
labeling according to size, we necessarily have $n_1=n_2$. We define
$n_*:=n_1=n_2$. From \eqref{E:mix_K} we have
\[%
   \partial_{k_1,k_2}^2\omega_{n_*}(k^{(1)})=0
\]%
and using \eqref{E:der2_rot_sym} with $\alpha=\pi/3$, we obtain
\begin{align*}
 \partial_{k_1}^2\omega_{n_*}(k^{(2)})
 &=\tfrac{1}{4}\big(\partial_{k_1}^2\omega_{n_*}(k^{(1)})
   +3\partial_{k_2}^2\omega_{n_*}(k^{(1)})\big),\\
 \partial_{k_2}^2\omega_{n_*}(k^{(2)})
 &=\tfrac{1}{4}\big(3\partial_{k_1}^2\omega_{n_*}(k^{(1)})
   +\partial_{k_2}^2\omega_{n_*}(k^{(1)})\big),\\
 \partial_{k_1,k_2}^2\omega_{n_*}(k^{(2)})
 &=\tfrac{\sqrt{3}}{4}\big(\partial_{k_1}^2\omega_{n_*}(k^{(1)})
   -\partial_{k_2}^2\omega_{n_*}(k^{(1)})\big).
\end{align*}
After having numerically checked the sets $\CA_{\alpha,\beta,\gamma,j}$ for all
combinations of $\alpha,\beta,\gamma,j$ to determine the nonlinear terms, we
thus arrive at the CMEs
\beq\label{E:CME_K_Kprime}%
 \begin{split}
  \big(\Omega +\alpha_1\partial_{y_1}^2
   +\beta_1\partial_{y_2}^2\big)A_1+\big(\gamma_0|A_1|^2
   +2\gamma_1|A_2|^2\big)A_1 &=0, \\
  \big(\Omega +\alpha_2\partial_{y_1}^2+\beta_2\partial_{y_2}^2
   +\mu\partial_{y_1,y_2}^2\big)A_2+\big(\gamma_0|A_2|^2
   +2\gamma_1|A_1|^2\big)A_2 &=0,
 \end{split}
\eeq%
where $\alpha_1=\tfrac{1}{2}\partial_{k_1}^2\omega_{n_*}(k^{(1)})$,
$\beta_1=\tfrac{1}{2}\partial_{k_2}^2\omega_{n_*}(k^{(1)})$, and
$\alpha_2=\tfrac{1}{4}(\alpha_1+3\beta_1)$,
$\beta_2=\tfrac{1}{4}(3\alpha_1+\beta_1)$,
$\mu=\tfrac{\sqrt{3}}{2}(\alpha_1-\beta_1)$, $\gamma_0:=
I_{1,1,1,1}=I_{2,2,2,2}$ using symmetry \eqref{E:I_sym_rot} with $\nu=\pi/3$,
and $\gamma_1:= I_{1,2,1,2}=I_{1,2,2,1}$ using \eqref{E:I_sym1}.

\paragraph{Example of Coupled Mode Equations for
$N=3$.}\label{S:CME_3_extrema}\

Let us assume that a gap edge for $N=3$ has extremal points at $k^{(1)}=M$,
$k^{(2)}=r_{\pi/3}(M)$, $k^{(3)}=r_{2\pi/3}(M)$. With the choice of the
reciprocal lattice vectors $b^{(1)}, b^{(2)}$ as in Section \ref{S:hex_struct}
we have $k^{(1)}=\tfrac{1}{2}b^{(2)}$,
$k^{(2)}=\tfrac{1}{2}\big(b^{(1)}+b^{(2)}\big)$, and
$k^{(3)}=\tfrac{1}{2}b^{(1)}$ with the corresponding integer shift sets $M_1 =
\{\bspm 0\\ 0 \espm, \bspm 0\\ 1 \espm\}$, $M_2 = \{\bspm 0 \\0 \espm, \bspm
1\\ 1 \espm\}$, and $M_3 = \{\bspm 0\\ 0 \espm, \bspm 1\\ 0 \espm\}$. Similarly
to Section \ref{S:CME_N2} we have $n_1=n_2=n_3=:n_*$. Using
\eqref{E:der2_rot_sym} and \eqref{E:mix_Mprime}--\eqref{E:M_der}, we get
\begin{align*}
   &\partial_{k_1}^2\omega_{n_*}(k^{(1)})
    =\partial_{k_1}^2\omega_{n_*}(k^{(2)})
    =\partial_{k_1}^2\omega_{n_*}(k^{(3)})
    =\partial_{k_2}^2\omega_{n_*}(k^{(1)})
    =\partial_{k_2}^2\omega_{n_*}(k^{(2)})
    =\partial_{k_2}^2\omega_{n_*}(k^{(3)})=:\alpha,\\
   &\partial_{k_1,k_2}^2\omega_{n_*}(k^{(1)})
    =\partial_{k_1,k_2}^2\omega_{n_*}(k^{(2)})
    =\partial_{k_1,k_2}^2\omega_{n_*}(k^{(3)})=0.
\end{align*}%
\begin{table}[h!]
\footnotesize
\begin{center}
\begin{tabular}{|c|c|c|c|c|c|c|}
\hline
$j$ & term & $\bspm\alpha \\ \beta\\ \gamma\espm$ & $k^{(\alpha)}+k^{(\beta)}$&
\multicolumn{2}{|c|}{$(n,o,q)^T$ from $\CA_{\alpha,\beta,\gamma,j}$ }&
coefficient of \\
 & in $\CN_j$ & & $-k^{(\gamma)}-k^{(j)}$ & $m=M_j(:,1)$ & $m=M_j(:,2)$ & the
term\\
\hline
1 & $|A_1|^2A_1$ & $\bspm 1\\1\\1\espm$ & $\bspm 0\\0\espm$ & $\bspm 0 & 0 \\0
&0\\0&0\espm, \bspm 0&0\\0&1\\0&1\espm$ & $\bspm 0&1\\0&0\\0&0\espm, \bspm
0&1\\0&1\\0&1\espm$ & $I_{1,1,1,1}$\\
\cline{2-7}
 & $|A_2|^2A_1$ & $\bspm 1\\2\\2\espm$ & $\bspm 0\\0\espm$ & $\bspm
0&0\\0&0\\0&0\espm, \bspm 0&0\\1&1\\1&1\espm$ & $\bspm 0&1\\0&0\\0&0\espm,
\bspm 0&1\\1&1\\1&1\espm$ & $2 I_{1,2,2,1}$\\
\cline{3-6}
 & & $\bspm 2\\1\\2\espm$ & $\bspm 0\\0\espm$ & $\bspm 0&0\\0&0\\0&0\espm,
\bspm 0&0\\1&1\\1&1\espm, \bspm 1&0\\0&1\\1&1\espm, \bspm 1&0\\-1&0\\0&0\espm$
& $\bspm 0&1\\0&0\\0&0\espm, \bspm 0&1\\1&1\\1&1\espm, \bspm
1&1\\0&1\\1&1\espm, \bspm 1&1\\-1&0\\0&0\espm$ & \\
\cline{2-7}
 & $|A_3|^2A_1$ & $\bspm 1\\3\\3\espm$ & $\bspm 0\\0\espm$ & $\bspm
0&0\\0&0\\0&0\espm, \bspm 0&0\\1&0\\1&0\espm$ & $\bspm 0&1\\0&0\\0&0\espm,
\bspm 0&1\\1&0\\1&0\espm$ & $2 I_{1,3,3,1}$\\
\cline{3-6}
& & $\bspm 3\\1\\3\espm$ & $\bspm 0\\0\espm$ & $\bspm 0&-1\\0&1\\0&0\espm,
\bspm 0&-1\\1&1\\1&0\espm, \bspm 1&0\\0&0\\1&0\espm, \bspm 1&0\\-1&0\\0&0\espm$
& $\bspm 0&0\\0&1\\0&0\espm, \bspm 0&0\\1&1\\1&0\espm, \bspm
1&1\\0&0\\1&0\espm, \bspm 1&1\\-1&0\\0&0\espm$ & \\
\cline{2-7}
 & $A_2^2\barl{A_1}$ & $\bspm 2\\2\\1\espm$ & $b^{(1)}$ & $\bspm
0&0\\1&0\\0&0\espm, \bspm 0&0\\1&1\\0&1\espm, \bspm 1&0\\0&0\\0&0\espm, \bspm
1&0\\0&1\\0&1\espm$ & $\bspm 0&1\\1&0\\0&0\espm, \bspm 0&1\\1&1\\0&1\espm,
\bspm 1&1\\0&0\\0&0\espm, \bspm 1&1\\0&1\\0&1\espm$ & $I_{2,2,1,1}$\\
\cline{2-7}
 & $A_3^2\barl{A_1}$ & $\bspm 3\\3\\1\espm$ & $b^{(1)}-b^{(2)}$ & $\bspm
0&-1\\1&0\\0&0\espm, \bspm 0&-1\\1&1\\0&1\espm, \bspm 1&0\\0&0\\0&1\espm, \bspm
1&0\\0&-1\\0&0\espm$ & $\bspm 0&0\\1&0\\0&0\espm, \bspm 0&0\\1&1\\0&1\espm,
\bspm 1&1\\0&0\\0&1\espm, \bspm 1&1\\0&-1\\0&0\espm$ & $I_{3,3,1,1}$\\
\hline
2 & $A_1^2\barl{A_2}$ & $\bspm 1\\1\\2\espm$ & $-b^{(1)}$ & $\bspm 0 & 0 \\0
&1\\1&1\espm, \bspm 0&0\\-1&0\\0&0\espm, \bspm -1&0\\0 &0\\0&0\espm, \bspm -1&0
\\1 &1\\1&1\espm$ & $\bspm 0&1\\0&0\\0&0\espm, \bspm 0&1\\1&1\\1&1\espm, \bspm
1&1\\0&1\\1&1\espm, \bspm 1&1\\-1&0\\0&0\espm$ & $I_{1,1,2,2}$\\
\cline{2-7}
 & $|A_1|^2A_2$ & $\bspm 1\\2\\1\espm$ & $\bspm 0\\0\espm$ & $\bspm
0&0\\0&0\\0&0\espm, \bspm 0&0\\0&1\\0&1\espm, \bspm -1&0\\1&0\\0&0\espm, \bspm
-1&0\\1&1\\0&1\espm$& $\bspm 0&1\\1&0\\0&0\espm, \bspm 0&1\\1&1\\0&1\espm, \bspm
1&1\\0&0\\0&0\espm, \bspm 1&1\\0&1\\0&1\espm $ & $2 I_{1,2,1,2}$\\
\cline{3-6}
& & $\bspm 2\\1\\1\espm$ & $\bspm 0\\0\espm$ & $\bspm 0&0\\0&0\\0&0\espm, \bspm
0&0\\0&1\\0&1\espm$ & $\bspm 1&1\\0&0\\0&0\espm, \bspm 1&1\\0&1\\0&1\espm$ &
\\
\cline{2-7}
 & $|A_2|^2A_2$ & $\bspm 2\\2\\2\espm$ & $\bspm 0\\0\espm$ & $\bspm
0&0\\0&0\\0&0\espm, \bspm 0&0\\1&1\\1&1\espm$ & $\bspm 1&1\\0&0\\0&0\espm,
\bspm 1&1\\1&1\\1&1\espm$ & $I_{2,2,2,2}$\\
\cline{2-7}
 & $|A_3|^2A_2$ & $\bspm 2\\3\\3\espm$ & $\bspm 0\\0\espm$ & $\bspm
0&0\\0&0\\0&0\espm, \bspm 0&0\\1&0\\1&0\espm$& $\bspm 1&1\\0&0\\0&0\espm, \bspm
1&1\\1&0\\1&0\espm$ & $2 I_{2,3,3,2}$\\
\cline{3-6}
 & & $\bspm 3\\2\\3\espm$ & $\bspm 0\\0\espm$ & $\bspm 0&0\\0&0\\0&0\espm,
\bspm 0&0\\1&0\\1&0\espm, \bspm 0&-1\\0&1\\0&0\espm, \bspm
0&-1\\1&1\\1&0\espm$& $\bspm 1&0\\0&1\\0&0\espm, \bspm 1&0\\1&1\\1&0\espm,
\bspm 1&1\\0&0\\0&0\espm, \bspm 1&1\\1&0\\1&0\espm $ & \\
\cline{2-7}
  & $A_3^2\barl{A_2}$ & $\bspm 3\\3\\2\espm$ & $-b^{(2)}$ & $\bspm
0&0\\1&0\\1&1\espm, \bspm 0&0\\0&-1\\0&0\espm, \bspm 0&-1\\0&0\\0&0\espm, \bspm
0&-1\\1&1\\1&1\espm$ & $\bspm 1&0\\0&0\\0&0\espm, \bspm 1&0\\1&1\\1&1\espm,
\bspm 1&1\\1&0\\1&1\espm, \bspm 1&1\\0&-1\\0&0\espm$ & $I_{3,3,2,2}$\\
\hline
3 & $A_1^2\barl{A_3}$ & $\bspm 1\\1\\3\espm$ & $b^{(2)}-b^{(1)}$ & $\bspm
0&1\\0&0\\1&0\espm, \bspm 0&1\\-1&0\\0&0\espm, \bspm -1&0\\0&1\\0&0\espm, \bspm
-1&0\\1&1\\1&0\espm$& $\bspm 0&0\\0&1\\0&0\espm, \bspm 0&0\\1&1\\1&0\espm,
\bspm 1&1\\0&0\\1&0\espm, \bspm 1&1\\-1&0\\0&0\espm$ & $I_{1,1,3,3}$\\
\cline{2-7}
& $|A_1|^2 A_3$ & $\bspm 1\\3\\1\espm$ & $\bspm 0\\0\espm$ & $\bspm
0&1\\0&0\\0&1\espm, \bspm 0&1\\0&-1\\0&0\espm, \bspm -1&0\\1&0\\0&0\espm, \bspm
-1&0\\1&1\\0&1\espm$& $\bspm 0&0\\1&0\\0&0\espm, \bspm 0&0\\1&1\\0&1\espm,
\bspm 1&1\\0&0\\0&1\espm, \bspm 1&1\\0&-1\\0&0\espm$ & $2I_{1,3,1,3}$\\
\cline{3-6}
& & $\bspm 3\\1\\1\espm$ & $\bspm 0\\0\espm$ & $\bspm 0&0\\0&0\\0&0\espm, \bspm
0&0\\0&1\\0&1\espm$& $\bspm 1&0\\0&0\\0&0\espm, \bspm 1&0\\0&1\\0&1\espm$ & \\
\cline{2-7}
& $A_2^2 \barl{A_3}$ & $\bspm 2\\2\\3\espm$ & $b^{(2)}$ & $\bspm
0&0\\0&1\\0&0\espm, \bspm 0&0\\1&1\\1&0\espm, \bspm 0&1\\0&0\\0&0\espm, \bspm
0&1\\1&0\\1&0\espm$& $\bspm 1&0\\0&1\\0&0\espm, \bspm 1&0\\1&1\\1&0\espm, \bspm
1&1\\0&0\\0&0\espm, \bspm 1&1\\1&0\\1&0\espm$ & $I_{2,2,3,3}$\\
\cline{2-7}
& $|A_2|^2 A_3$ & $\bspm 2\\3\\2\espm$ & $\bspm 0\\0\espm$ & $\bspm
0&0\\0&0\\0&0\espm, \bspm 0&0\\1&1\\1&1\espm, \bspm 0&1\\1&0\\1&1\espm, \bspm
0&1\\0&-1\\0&0\espm$& $\bspm 1&0\\0&0\\0&0\espm, \bspm 1&0\\1&1\\1&1\espm,
\bspm 1&1\\1&0\\1&1\espm, \bspm 1&1\\0&-1\\0&0\espm$ & $2I_{2,3,2,3}$\\
\cline{3-6}
& & $\bspm 3\\2\\2\espm$ & $\bspm 0\\0\espm$ & $\bspm 0&0\\0&0\\0&0\espm, \bspm
0&0\\1&1\\1&1\espm$& $\bspm 1&0\\0&0\\0&0\espm, \bspm 1&0\\1&1\\1&1\espm$ & \\
\cline{2-7}
 & $|A_3|^2A_3$ & $\bspm 3\\3\\3\espm$ & $\bspm 0\\0\espm$ & $\bspm
0&0\\0&0\\0&0\espm, \bspm 0&0\\1&0\\1&0\espm$& $\bspm 1&0\\0&0\\0&0\espm, \bspm
1&0\\1&0\\1&0\espm$ & $I_{3,3,3,3}$\\
\hline
\end{tabular}
   \caption{\label{T:NL_terms_3extrema}\small%
Calculation of the nonlinear terms for Section \ref{S:CME_3_extrema}.}
\end{center}
\end{table}
The sets $\CA_{\alpha,\beta,\gamma,j}$ are, once again, determined using the
Matlab routine and the results are for illustration listed in Table
\ref{T:NL_terms_3extrema}. The resulting CMEs are
\beq\label{E:CME_3extrema}%
 \begin{split}
  \left(\Omega +\alpha(\partial_{y_1}^2+\partial_{y_2}^2)\right)A_1
   +\left(\gamma_0|A_1|^2+2\gamma_1(|A_2|^2+|A_3|^2)\right)A_1
   +\gamma_2(A_2^2+A_3^2)\barl{A_1} & = 0,\\
  \left(\Omega+\alpha(\partial_{y_1}^2+\partial_{y_2}^2)\right)A_2
   +\left(\gamma_0|A_2|^2+2\gamma_1(|A_1|^2+|A_3|^2)\right)A_2
   +\gamma_2(A_1^2+A_3^2)\barl{A_2} & = 0,\\
  \left(\Omega +\alpha(\partial_{y_1}^2+\partial_{y_2}^2)\right)A_3
   +\left(\gamma_0|A_3|^2+2\gamma_1(|A_1|^2 +|A_2|^2)\right)A_3
   +\gamma_2(A_1^2+A_2^2)\barl{A_3} & = 0,
 \end{split}
\eeq%
where the following symmetries have been used:
$\gamma_0:=I_{1,1,1,1}=I_{2,2,2,2}=I_{3,3,3,3}$ due to \eqref{E:I_sym_rot} with
$\nu=\pi/3$; $\gamma_1:=I_{1,2,2,1}=I_{2,3,3,2}=I_{1,2,1,2}=I_{2,3,2,3}$ due to
\eqref{E:I_sym_rot} with $\nu=\pi/3$, and \eqref{E:I_sym1}. Moreover,
$\gamma_1=I_{2,1,1,2}=I_{1,3,3,1}=I_{3,1,1,3}$, where the second equality
follows from \eqref{E:I_sym_rot} with $\nu=\pi/3$ and the facts that
$k^{(1)}=r_{\pi/3}(k^{(3)}-b^{(1)})$ and
$u_n(k^{(3)}-b^{(1)};x)=u_n(k^{(3)};x)$ for all $n\in \N$. Finally
$\gamma_2:=I_{2,2,1,1}=I_{3,3,2,2}=\overline{I_{1,1,2,2}}=\overline{I_{2,2,3,3}}$
due to \eqref{E:I_sym_rot} and \eqref{E:I_sym2}, and
$\gamma_2=I_{2,2,1,1}=I_{1,1,3,3}$ using \eqref{E:I_sym_rot} together with
$k^{(1)}=r_{\pi/3}(k^{(3)}-b^{(1)})$ and
$u_n(k^{(3)}-b^{(1)};x)=u_n(k^{(3)};x)$ for all $n\in \N$. All the nonlinear
coefficients are real: $\gamma_0,\gamma_1\in \R$ due to \eqref{E:I_sym2} and
$\gamma_2\in \R$ since $\gamma_2=I_{2,2,1,1}=\overline{I_{1,1,2,2}}$ by
\eqref{E:I_sym2} and at the same time $\gamma_2=I_{2,2,1,1}=I_{1,1,2,2}$ by
\eqref{E:I_sym_refl}, where we are using the facts that
$k^{(2)}=(k^{(1)}_1,-k^{(1)}_2)^T$ and that $k^{(2)}\doteq k^{(1)}$ does not
hold.

\section{Numerical Examples of Gap Soliton Approximations}\label{S:numerics}

We compute here numerically localized solutions of the CMEs for the examples
$s_2,s_5$ in Section \ref{S:CME_hex}. Then, using the leading order term in
\eqref{E:ansatz_phys}, we generate and plot an approximation of a gap soliton
of the nonlinear Maxwell problem \eqref{E:NL_Maxw}. In the evaluation of
\eqref{E:ansatz_phys} we position the photonic crystal so that the center of
one of the annuli lies at the origin $x=0$.

\subsection{Gap Soliton near the Edge $s_2$}\label{S:numerics_s2}

\begin{figure}[h!]
\begin{center}
\epsfig{figure = 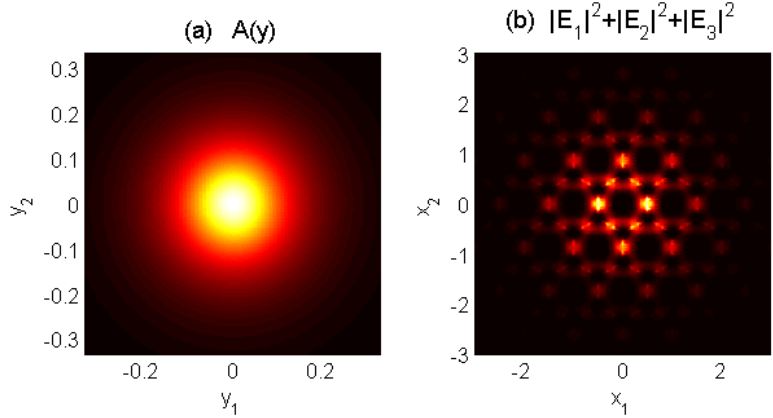,scale=0.65}
\caption{\label{F:s2_envel_GS} \small (a) CME solution,
(b) intensity of the gap soliton approximation for the case $\omega_*=s_2$.
See Section \ref{S:numerics_s2}.}
\end{center}
\end{figure}

\begin{figure}[h!]
\begin{center}
\hspace{-0.5cm}
\epsfig{figure = 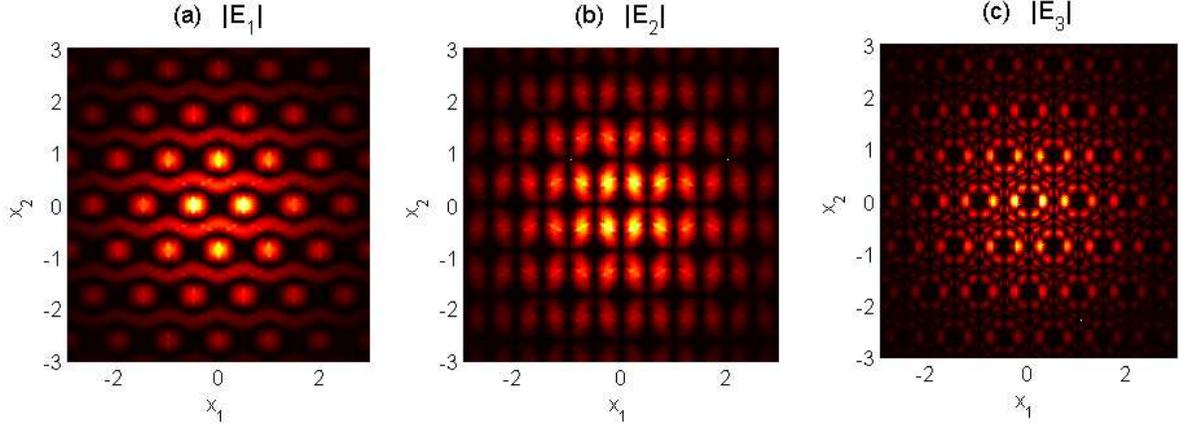,scale=0.65}
\caption{\label{F:s2_E1-E3}\small Absolute value of the components
$E_1,E_2,E_3$ of the gap soliton approximation for $\omega_*=s_2$.
See Section \ref{S:numerics_s2}.}
\end{center}
\end{figure}

Figure \ref{F:s2_envel_GS} plots in (a) the unique positive localized solution,
the so called Townes soliton, of \eqref{E:CME_1} for the case $\omega_*=s_2$
and in (b) the intensity $I=|E_1|^2+|E_2|^2+|E_3|^2$ of the leading order term
in \eqref{E:ansatz_phys}. In Figure \ref{F:s2_E1-E3} we show the absolute value
of the individual components $E_1, E_2, E_3$. As the Townes soliton is radially
symmetric, it was computed using the shooting method on \eqref{E:CME_1} in
polar coordinates. The fourth to fifth order explicit Runge--Kutta method ODE45
of Matlab was used in the shooting method.


\subsection{Gap Soliton near the Edge $s_5$}\label{S:numerics_s5}

\begin{figure}[h!]
\begin{center}
\epsfig{figure = 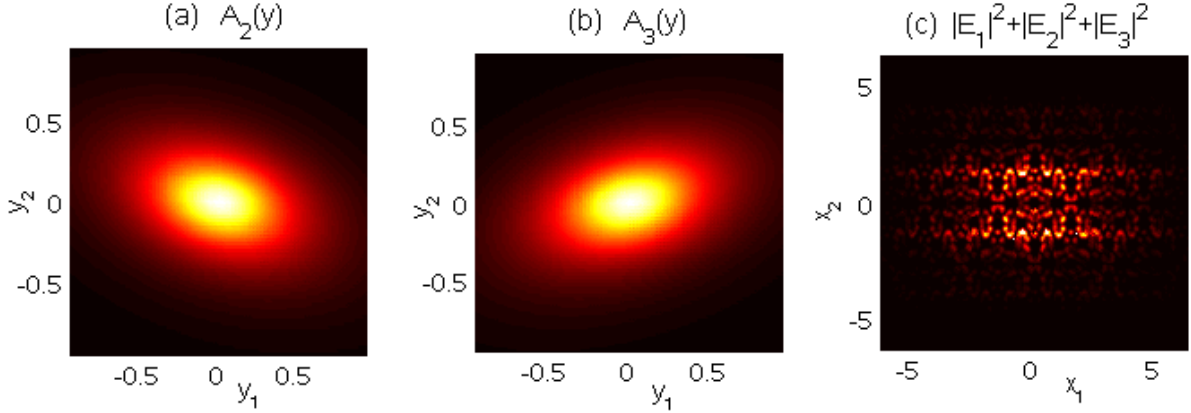,scale=0.9}
\end{center}
   \caption{\label{F:s5_envel_GS}\small (a) CME solutions $A_2, A_3$,
(b) intensity of the gap soliton approximation for the case $\omega_*=s_5$.
See Section \ref{S:numerics_s5}.}
\end{figure}

\begin{figure}[h!]
\begin{center}
\hspace{-0.5cm}
\epsfig{figure = 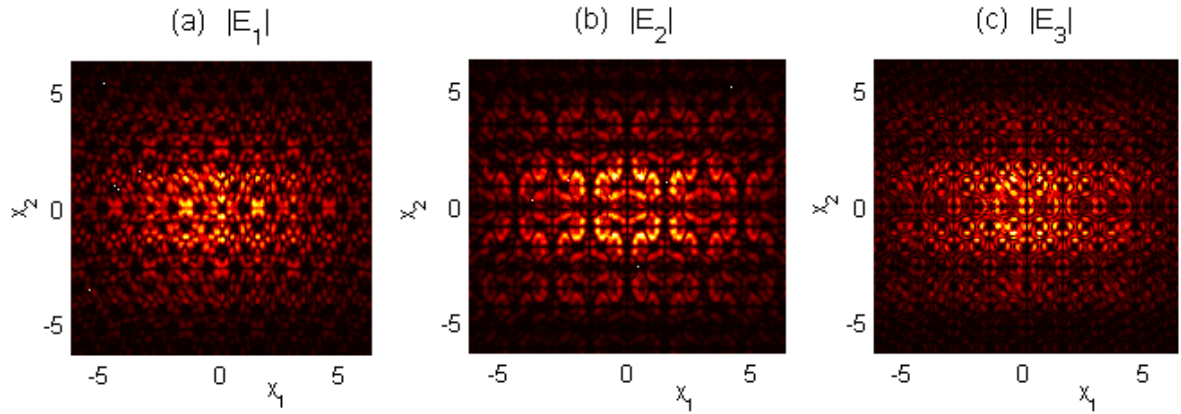,scale=0.9}
\caption{\label{F:s5_E1-E3}\small Absolute value of the components $E_1,E_2,E_3$
of the gap soliton approximation for $\omega_*=s_2$. See Section \ref{S:numerics_s5}.}
\end{center}
\end{figure}

Here we restrict to solutions of \eqref{E:CME_s5} with the symmetry
\[
   A_1=A_4,\ A_2=A_5,\ A_3=A_6,
\]%
which reduces the problem to a system of three equations for $A_1,A_2,A_3$. To
find a localized solution, we first replace $\mu$ by $0$, and
$\alpha_1,\alpha_2,\beta_1,\beta_2$ by the average of these four numbers. Also
the coefficients in each $\CN_j$, $j\in \{1,2,3\}$, are replaced by their
average. For this modified system the Townes soliton with $A_1=A_2=A_3$ is
computed via the shooting method as in Section \ref{S:numerics_s2}. Then a
numerical homotopy in the coefficients is used to get a solution of
\eqref{E:CME_s5}. The homotopy is applied to a fourth order centered finite
difference discretization of \eqref{E:CME_s5}. Our homotopy always results in
$A_1=0$ so that in the end we produce a solution of \eqref{E:CME_s5} with
$A_1=A_4=0$ and $A_2=A_5\neq 0$, $A_3=A_6\neq 0$. The two components $A_2,A_3$
are plotted in Figure \ref{F:s5_envel_GS} together with the intensity of the
corresponding leading order term in \eqref{E:ansatz_phys}. In Figure
\ref{F:s5_E1-E3} we plot the individual components of $E$ in absolute value.

\section{Conclusions}
We have considered monochromatic out-of-plane gap solitons in Kerr nonlinear 2D
photonic crystals as described by the full vector Maxwell system. Using a model
of the nonlinear polarization which does not produce higher harmonics, we
arrive at a cubically nonlinear curl-curl problem for the fundamental harmonic.
For gap solitons with frequencies in spectral gaps but in an asymptotic
vicinity of a gap edge we assume a standard slowly varying envelope
approximation based on the gap edge Bloch waves modulated by slowly varying
envelopes of small amplitude. These envelopes are then shown to satisfy a
system of coupled mode equations (CMEs) of the same structure as in the case of
gap solitons of the 2D periodic nonlinear Schr\"odinger equation
\cite{DU09,DU_err11}. In particular the system generally involves mixed
derivatives. Being a constant coefficient system depending only on the slow
variables, the CMEs is a simple effective model for the near edge gap solitons.
Similarly to \cite{DU09} the derivation of CMEs needs to be carried out in
Bloch variables due to the possible quasi-periodicity of gap edge Bloch waves.
Symmetries among the coefficients of the CMEs are determined using symmetries
of the band structure and among the Bloch waves.

We provide an example of a photonic crystal with a hexagonal periodicity
lattice and a circular material structure in the periodicity cell. For this
crystal three gaps are numerically observed (for $\omega>0$). CMEs are then
derived for several gap edges including a case where a system of six CMEs
arises. Numerical computations of localized solutions of these CMEs and of the
corresponding gap soliton approximations are then performed. For the CME system
with six components only solutions with four nonzero components were
numerically constructed and it is unclear whether a solution with all six
nonzero components exists.

A rigorous justification of the CMEs, which states that for a certain class of
CME solutions the full Maxwell system has gap soliton solutions which are
indeed approximated by the slowly varying envelope asymptotic expansion, is
expected to hold by similar arguments to those in \cite{DPS09,DU09,DU_err11}.
It will be the subject of future work.

\section*{Acknowledgments}
We thank Stefan Findeisen, Karlsruhe Institute of Technology, for carrying out
the finite element computations in Section \ref{S:hex_struct}. T. Dohnal was
partially supported by DFG Research Training Group 1924: Analysis, Simulation
and Design of Nanotechnological Processes.

\bibliographystyle{plain}
\bibliography{bibliography}

\end{document}